\documentclass[lettersize,journal]{IEEEtran}
\usepackage{amsmath,amsfonts}
\usepackage{algorithmic}
\usepackage{algorithm}
\usepackage{array}
\usepackage[caption=false,font=normalsize,labelfont=sf,textfont=sf]{subfig}
\usepackage{textcomp}
\usepackage{stfloats}
\usepackage{url}
\usepackage{verbatim}
\usepackage{graphicx}
\usepackage{cite}
\usepackage{enumitem} 
\usepackage{hyperref}
\usepackage{multirow}
\usepackage{booktabs} 
\usepackage{colortbl}
\usepackage{xcolor}         
\usepackage{wrapfig}
\usepackage{balance}

\def\model{Ghost }
\def\modela{SKT }
\def\modelb{AUO }
\def\base1{LETTER }
\def\base2{LC-Rec }
\def\base3{ED$^2$ }
\def\repolink{\href{https://github.com/Esperanto-mega/Ghost}{\textit{Ghost Project}}}

\newcommand{\third}{\cellcolor[HTML]{E6F2FF}} 
\newcommand{\first}{\cellcolor[HTML]{B2DFDB}}  
\newcommand{\second}{\cellcolor[HTML]{C5CAE9}} 

\begin{document}

\title{Echoes in Filter Bubble: Diagnosing and Curing Popularity Bias in Generative Recommenders}

\author{
\IEEEauthorblockN{
Jun Yin\IEEEauthorrefmark{1},
Bangguo Zhu\IEEEauthorrefmark{1},
Peng Huo,
Ruochen Liu,
Hao Chen,
Senzhang Wang,\\
Shirui Pan\IEEEauthorrefmark{2},~\IEEEmembership{Senior Member,~IEEE}
Chengqi Zhang\IEEEauthorrefmark{2},~\IEEEmembership{Fellow,~IEEE}
}
\IEEEcompsocitemizethanks{
\IEEEcompsocthanksitem Jun Yin and Chengqi Zhang are with the Department of Data Science and Artificial Intelligence, Hong Kong Polytechnic University, Hong Kong SAR, China. Email: Junmay.yin@connect.polyu.hk, Chengqi.zhang@polyu.edu.hk
\IEEEcompsocthanksitem Bangguo Zhu, Ruochen Liu, and Senzhang Wang are with the School of Computer Science and Engineering, Central South University, Changsha, China. Email: \{8210231122, ruochen, szwang\}@csu.edu.cn
\IEEEcompsocthanksitem Peng Huo is with the National Super Computing Center, Tianjin, China. Email: huopeng@nscc-tj.cn
\IEEEcompsocthanksitem Hao Chen is with the Faculty of Data Science, City University of Macau, Macau, China. Email: haochen@cityu.edu.mo
\IEEEcompsocthanksitem Shirui Pan is with the School of Information and Communication Technology, Griffith University, Brisbane, Australia. Email: s.pan@griffith.edu.au
}
\thanks{* Equal Contribution. \dag~Corresponding Authors.}
}


\markboth{
Preprint Manuscript
}%
{Shell \MakeLowercase{\textit{et al.}}: A Sample Article Using IEEEtran.cls for IEEE Journals}


\maketitle

\begin{abstract}
Recently, Generative Recommenders (GRs), characterized by a unified end-to-end framework, have exhibited astonishing potential in transforming the recommendation paradigm. Despite their effectiveness, we recognize that GRs are still susceptible to the long-standing issue of popularity bias that has pervaded the recommendation community. Although a few studies have attempted to extend traditional debiasing methods to GRs, their effectiveness is marginal, and the fundamental reason why GRs suffer from popularity bias remains under-explored. To bridge this gap, this study focuses on two core aspects in GRs: the optimization of generative framework and the item tokenization based on semantic index. Based on theoretical analyses, we identify that the severe popularity bias emerges from the confluence of a token-level optimization flaw and the undifferentiated property of item tokenization. Accordingly, this study develops a novel generative recommender system, called Ghost\footnote{The Ghost project is available at \repolink.}, by designing the asymmetric unlikelihood optimization and the skeleton-founded tokenization. Extensive empirical evaluations across three datasets, alongside multiple SOTA baselines, reveal that \model substantially alleviates popularity bias and promotes fairer recommendations, while incurring slight degradation to the overall recommendation utility.
\end{abstract}


\begin{figure*}[t]
    \centering
    \includegraphics[width=\textwidth]{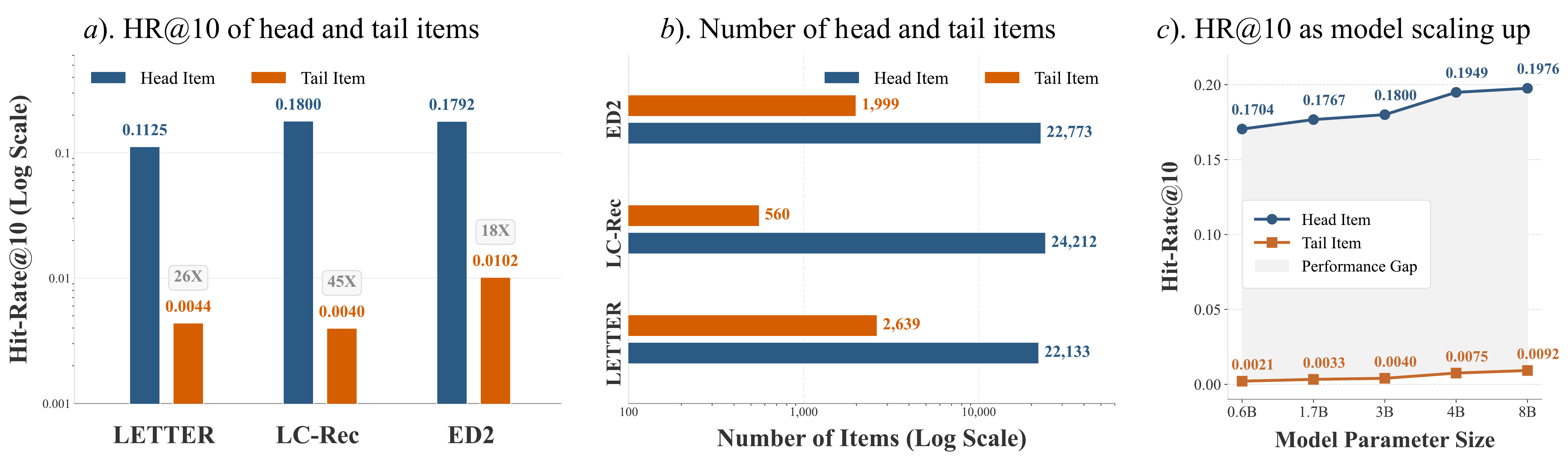}
    \caption{
    \textit{a}). Comparison of Hit-Rate@10 (i.e., HR@10) between head and tail items. \textit{b}). Comparison between the number of head and tail items in the recommendation list provided by three GRs. \textit{c}). Tendency of HR@10 as the backbone parameters of LC-Rec scaling up.}
    \label{fig:intro}
    \vspace{-0.4cm}
\end{figure*}

\begin{figure}[t]
    \centering
    \includegraphics[width=\linewidth]{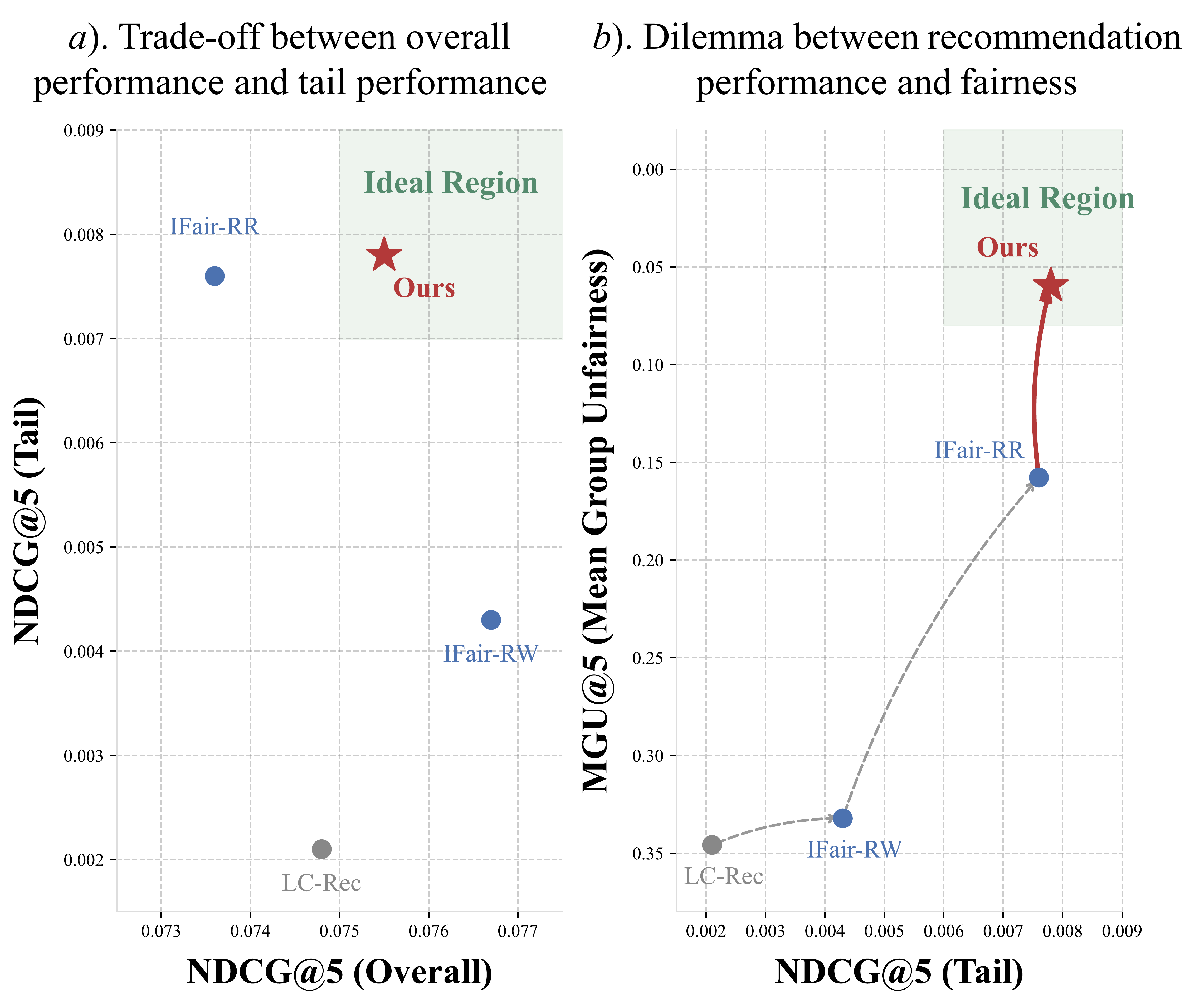}
    \caption{Limitations of current popularity debiasing methods on GRs. \textit{a}). Trade-off between overall recommendation performance and that of tail items. \textit{b}). Dilemma between performance and fairness of recommendation results.}
    \label{fig:trade_off}
\vspace{-0.2cm}
\end{figure}

\section{Introduction}
\IEEEPARstart{R}{ecommender} systems, due to the ability to capture user preferences by analyzing historical interactions, are essential for enhancing user experiences across platforms such as e-commerce \cite{alibaba}, video streaming \cite{youtube}, and social networks \cite{deepinf}. Recently, generative recommenders (GRs) have emerged as a transformative paradigm \cite{tiger,lcrec,letter,ed2,mql4grec} by replacing traditional item IDs \cite{lightgcn,sasrec} with semantic indices (SIDs) \cite{vqvae,rqvae}. Driven by the adoption of large language models (LLMs) \cite{deepseek,llama,GPT} as the backbone, the architecture of state-of-the-art (SOTA) GRs appears to become increasingly well-established, ushering in an era of rapid advancement.

However, despite their significant improvements in recommendation performance, we identify that current leading GRs remain plagued by the persistent problem of popularity bias \cite{wsdm25bestpaper,eliminate_popularity_bias,popularity_opportunity_bias}, which has long affected the recommender systems field \cite{ifair}. As illustrated in Figure \ref{fig:intro}, three state-of-the-art (SOTA) GRs, LETTER \cite{letter}, LC-Rec \cite{lcrec}, and ED$^2$ \cite{ed2}, all exhibit an extreme over-recommendation phenomenon for popular head items while struggling to accurately predict the niche tail items. In particular, Figure \ref{fig:intro}\textit{a}) showcases the significant gap between the performance of head items and tail items, which is up to \textit{45 times} greater for LC-Rec. Figure \ref{fig:intro}\textit{b}) presents the number of head and tail items in recommendation lists and reveals an obvious preference for popular items, which account for \textit{more than 97 percent} of the recommendation lists. Furthermore, according to Figure \ref{fig:intro}\textit{c}), as the backbone scale of LC-Rec increases \textit{from 0.6B to 8B}, the popularity bias is not mitigated spontaneously. Consequently, the popularity bias in GRs precipitates a filter bubble, wherein trending content monopolizes visibility while high-quality, long-tail items are severely marginalized. 

Although a few preliminary studies \cite{ifair} have attempted to extend conventional methods of popularity debiasing, such as item re-weighting \cite{reweight_1,reweight_2} and result re-ranking \cite{rerank_1,rerank_2}, to GRs, their effectiveness is relatively marginal. As illustrated by Figure \ref{fig:trade_off}, existing methods (i.e., IFair-RW and IFair-RR) \cite{ifair} still struggle to achieve appropriate \textit{Pareto Optimality} \cite{pareto_1,pareto_2} while collectively considering the overall recommendation performance, tail item recommendation performance, and recommendation fairness. A non-negligible distance persists from the ideal region. On the other hand, the fundamental reason why GRs suffer from severe popularity bias is still under-explored. Compared with the discriminative paradigm \cite{ncf,sasrec,lightgcn}, two distinctive characteristics of GRs are \textit{(i)} the end-to-end generative framework and \textit{(ii)} the item tokenization based on SIDs. However, these two distinct aspects are completely overlooked when designing popularity debiasing methods for GRs, rendering them devoid of strategic principles.

To bridge the current gap, this study is anchored in the optimization dynamics of the generative framework and the SID structures under current tokenization. \textit{Firstly}, regarding the optimization based on maximum likelihood estimation (MLE), which is widely adopted by current GRs \cite{tiger,lcrec,ed2}, we start with a gradient analysis and identify that the SID tokens of tail items mostly suffer from a gradient starvation issue. Due to the heavily long-tailed training distribution \cite{wsdm25bestpaper,2_8_split}, the tail item tokens are pathologically pushed away from user preference. Hence, during the recommendation process, when head and tail items compete, the optimization flaw leads to the dominance of head item tokens. \textit{Secondly}, this study sheds light on the fact that the current item tokenization is undifferentiated for head and tail items, without accounting for disparities in item popularity. It induces unpredictable item competition and continuously amplifies the token-level bias between head and tail items. Eventually, the undifferentiated tokenization renders the probability of tail items severely hijacked by their popular counterparts, and thus the whole GR model struggles to cast off the popularity bias.

Based on the crucial insights above, we develop a novel generative recommender named Ghost \footnote{\textit{Ghost} denotes a \textbf{G}enerative recommender with asymmetric unlikeli\textbf{h}ood \textbf{o}ptimization and \textbf{s}keleton-founded \textbf{t}okenization.}, which is equipped with the asymmetric unlikelihood optimization (AUO) and the skeleton-founded tokenization (SKT). In detail, to mitigate the gradient starvation issue, \modelb introduces asymmetric token-level unlikelihood between the tail items and their head counterparts. By constructing a reasonable undesired collection, \modelb succeeds in rescuing the ineffective gradients of tail item tokens. To inhibit the bias amplification effect caused by undifferentiated tokenization, \modela first designates the head item SIDs as the skeleton of the whole SID system. Subsequently, tail item SIDs are cultivated along the skeleton structure, with a dedicated focus on capturing the distinctiveness between tail and head items. The main contributions of this study are summarized as follows.

\begin{itemize}
[leftmargin=*]

\item This study diagnoses the troublesome issue of popularity bias in generative recommenders and identifies two fundamental factors, \textit{(i)} the gradient starvation problem of tail item tokens and \textit{(ii)} the bias amplification effect of undifferentiated item tokenization.

\item This study develops Ghost, a novel GR that exhibits resilience to popularity bias, by designing the asymmetric unlikelihood optimization (AUO) and the skeleton-founded item tokenization (SKT). Specifically, \modelb integrates effective negative feedback to calibrate the optimization of tail item tokens. Based on the skeleton structure, \modela ingeniously reduces disordered head-tail competition and characterizes the distinctiveness of tail items.

\item Extensive empirical evaluations across three datasets, alongside multiple SOTA baselines, reveal that \model substantially alleviates popularity bias and promotes fairer recommendations, ultimately achieving the desired Pareto optimality of generative recommenders.
\end{itemize}


\section{Preliminary}\label{sec:preliminary}
\textbf{Problem Formulation}. This study focuses on the \textit{sequential recommendation} task \cite{sasrec,lcrec,ed2}, which aims to predict the next most suitable item based on the user historical behavior. Considering a system of $K$ items $\{v_k|k=1,2,\cdots,K\}$ and $J$ users $\{u_j|j=1,2,\cdots,J\}$, the behavior of $u_j$ can be represented by an item sequence $h_{u_j} = [v_{k_1},v_{k_2},\cdots,v_{k_l}]$, where $l$ is the sequence length.

\textbf{Generative Recommender Systems}. Existing efforts towards developing GRs \cite{tiger,lcrec,letter,ed2,care} primarily concentrate on the paradigm based on SIDs. Generally, SID-based GRs consist of two main components, i.e., item tokenization and recommendation-oriented finetuning \cite{lcrec,ed2}. For item tokenization, SID-based GRs usually introduce vector quantization techniques, such as VQ-VAE \cite{vqvae}, RQ-VAE \cite{rqvae}, and RQ-KMeans \cite{rqkmeans}, to convert the continuous embeddings of items into discrete indices. Taking RQ-VAE as an example, and assuming that $X_v$ is the item embedding and the collection $\{\mu^{(i)}_n\}_{n=1}^N$ denotes the $i$-th codebook within the RQ-VAE, the SID generation process can be represented as,
\begin{equation}
\label{eq:head_item_sid}
\begin{split}
    c_{v}^{(i)} &= \arg\min_{n\in\{1, 2,\dots,N\}} \|{r}_{v}^{(i)}-{\mu}_{n}^{(i)}\|_{2}^{2}, \\
    {r}_{v}^{(i+1)} &= {r}_{v}^{(i)} - {\mu}_{c_{v}^{(i)}}^{(i)}, \quad \text{for } i=1,2,\dots,L.
\end{split}
\end{equation}
where $r_v^{(1)}=X_v$ and $L$ denotes the SID length. Afterwards, item $v$ can be indexed as an SID $\Omega_v=(c_v^{(1)},c_v^{(2)},\cdots,c_v^{(L)})$. For recommendation-oriented finetuning, the sequential recommendation task can be reformulated as a language generation task based on SIDs. In particular, for historical behavior $h_u$ and target item $v$ with SID $\Omega_v$, the optimization objective of SID-based GRs follows Maximum Likelihood Estimation (MLE) and usually adopts the negative log-likelihood (NLL) loss, as follows,
\begin{equation}
    \mathcal L_{\text{NLL}}=-\sum\nolimits_{i}\log \mathcal P_{\theta}(c_v^{(i)}|h_{u},c_v^{<i}),\label{eq:mle}
\end{equation}
where $\theta$ denotes the GR parameters and $c_v^{<i}$ denotes the sub-sequence of SID $\Omega_v$ before the $i$-th position. \textit{This study focuses on SID-based GRs, as they are currently the most influential paradigm with the greatest performance potential.}

\textbf{Popularity Bias}. In the domain of recommender systems, the interaction frequency of items often follows a long-tail distribution \cite{popularity_opportunity_bias,eliminate_popularity_bias,wsdm25bestpaper,mitigating_popularity_bias}, especially a power-law distribution. Popularity bias refers to the algorithmic preference to disproportionately favor frequently interacted short-head items at the expense of highly relevant but sparse long-tail items \cite{wsdm25bestpaper}. \textit{This study follows a common practice \cite{2_8_split} in popularity debiasing research, which groups the items into the head set (i.e., the top 20\% most popular items) and the tail set (i.e., the remaining 80\% of items) based on item popularity.}

\textbf{Pareto Optimality}. In multi-objective optimization, a model is simultaneously evaluated across $M$ distinct, often conflicting metrics  \cite{pareto_1}. Assuming that all objectives $\{f_m\}_{m=1}^M$ are to be minimized, a solution $s$ is said to strictly precede another solution $s'$, denoted as $s \prec s'$, if and only if $\forall ~m_1,f_{m_1}(s)\le f_{m_1}(s')$ and $\exists ~m_2, f_{m_2}(s)<f_{m_2}(s')$. A solution $s^*$ is defined as Pareto optimal if there exists no other feasible solution $s$ such that $s \prec s^*$. Consequently, Pareto optimality represents the optimal trade-off among the multiple evaluation metrics, indicating that no single objective can be further improved without fundamentally degrading at least one other objective \cite{pareto_1,pareto_2}.

\section{Diagnose Popularity Bias in GRs}\label{sec:diagnose}
To investigate the popularity bias in current GRs, this study starts with an analysis of \textit{(i)} the gradient of MLE optimization and \textit{(ii)} the SID structure under undifferentiated tokenization. 

\subsection{Gradient Analysis of MLE Optimization}\label{sec:mle_opt}
Under the MLE objective $\mathcal L_{\text{NLL}}$ defined in Eq.\eqref{eq:mle}, the optimization conditioned on user historical behavior $h_u$ is governed by the Softmax derivative. Let $\mathcal D$ denote the training distribution of user-item interaction pairs $(h_u,v)$, where the history $h_u$ is encoded into a representation $X_{h_u}$ and the target item $v$ is tokenized as a SID $\Omega_v=(c_v^{(1)},c_v^{(2)},\cdots,c_v^{(L)})$. 

\textbf{LEMMA 1 (Gradient Starvation in MLE)}. \textit{For an arbitrary SID token $c$, whose embedding is $e_c$, the expected MLE gradient update} $\mathbb E_\mathcal D[\Delta e_c]$ \textit{over the training distribution $\mathcal D$ conforms to}
\begin{equation}
\begin{split}
    \mathbb E_{\mathcal D}[\Delta e_c] &\propto \mathbb E_{\mathcal D}\Bigg[\sum_{i}\bigg(\mathbb I\{c=c_v^{(i)}\}\cdot(1-\mathcal P_\theta(c|h_u,c_v^{<i})) \\
    &\quad - \mathbb I\{c\neq c_v^{(i)}\}\cdot\mathcal P_\theta(c|h_u,c_v^{<i})\bigg)\cdot X_{h_u}\Bigg].
\end{split}
\end{equation}
\textit{However, for a tail token $c_{\text{tail}}$ that predominantly composes tail items, the expected gradient update is heavily skewed in the negative direction, formulated as,}
\begin{equation}
\begin{split}
    &\mathbb E_\mathcal D[\langle\Delta e_{c_{\text{tail}}},X_{h_u}\rangle] 
    \\ & \approx -\mathbb E_\mathcal D\left[\sum_{i}\mathcal P_\theta(c_{\text{tail}}|h_u,c_v^{<i})\cdot \Vert X_{h_u}\Vert_2^2\right] \leq 0.
\end{split}
\end{equation}

In a long-tailed distribution $\mathcal{D}$, tokens composing head items frequently act as the targets, receiving massive positive gradient updates that align the embedding $e_c$ with user preference $X_{h_u}$. Conversely, tokens that are specific to tail items mainly act as trivial negative samples in the denominator of the Softmax operation. This leads to the \textit{Gradient Starvation} issue \cite{gradient_starvation} for tail item tokens, in which the token embeddings are consistently pushed away from the user preference space.

\subsection{SID Branching Points under Undifferentiated Tokenization}
Subsequently, we further investigate how this token-level optimization flaw propagates into item-level bias during the recommendation process. Let $c_{\text{head}}^{(i)}$ and $c_{\text{tail}}^{(i)}$ be candidate tokens competing at the $i$-th item generation step, and step $i$ is denoted as a branching point. Due to the asymptotically repulsive gradient updates stemming from gradient starvation, the generation probability of the tail token $c_{\text{tail}}^{(i)}$ is systematically penalized, resulting in a bias amplification factor.

\textbf{COROLLARY 1 (Head Token Dominance at Branching Point).} \textit{At step $i$, where head and tail paths compete, the token generation probability ratio diverges from the true data distribution $\mathcal{P}_d$ by a local amplification factor $\gamma_i > 1$:}
\begin{equation}
\frac{\mathcal{P}_{\theta}(c_{\text{head}}^{(i)}|h_u, c^{<i})}{\mathcal{P}_{\theta}(c_{\text{tail}}^{(i)}|h_u, c^{<i})} = \gamma_i \cdot \frac{\mathcal{P}_d(c_{\text{head}}^{(i)}|h_u, c^{<i})}{\mathcal{P}_d(c_{\text{tail}}^{(i)}|h_u, c^{<i})}.\label{eq:dominance}
\end{equation}

Eq.\eqref{eq:dominance} implies that the generation process becomes pathologically overconfident in head tokens, regardless of the current context (i.e., $h_u$ and $c^{(<i)}$). Furthermore, current GRs mostly tokenize items into SIDs in an undifferentiated style, which treats items identically without accounting for their inherent popularity disparities. The undifferentiated nature induces that there is no predictable structural branching point between head and tail items. Within the undifferentiated tokenization, a tail item $v_{\text{tail}}$ shares prefixes of varying lengths with numerous popular items and is bound to encounter a sequence of branching points with head token dominance.

\textbf{LEMMA 2 (Bias Amplification via Undifferentiated Tokenization).} \textit{Let $\mathcal{Z}$ (where $|\mathcal{Z}|=z\leq L$) be the set of steps along the generation process of $v_{\text{tail}}$, where it must compete against head candidate tokens. The probability of successfully navigating these intersections without being hijacked by head items is geometrically suppressed during recommendation,}
\begin{equation}
\label{eq:lemma2}
\begin{split}
    \mathcal{P}_{\theta}(v_{\text{tail}}|h_u) 
    &= \prod_{j} \mathcal{P}_{\theta}(c_{\text{tail}}^{(j)}|h_u, c^{<j}) \\
    &\leq (\gamma_{\text{min}})^{-z} \cdot \prod_{j} \mathcal{P}_d(c_{\text{tail}}^{(j)}|h_u, c^{<j}).
\end{split}
\end{equation}
\textit{where $\gamma_{\text{min}} = \min_{j \in \mathcal{Z}}(\gamma_j)>1$. Therefore, the probability ratio of generating a competing head item over the tail item $v_{\text{tail}}$ cascades geometrically by a factor of at least $\mathcal{O}(\gamma_{\text{min}}^z)$.}

\textbf{Discussion.} 
According to the analyses above, the severe popularity bias emerges from the confluence of \textit{(i)} gradient starvation in MLE optimization and \textit{(ii)} bias amplification of undifferentiated tokenization. Detailed derivations are presented in Appendix \ref{app:theory}. Intuitively, under MLE optimization, tail item tokens fail to receive effective gradient signals, trapping them in gradient starvation. Then, the gradient starvation of tail item tokens results in the head token dominance at branching point. Finally, during the recommendation process, a sequence of head-dominated branching points are brought by the undifferentiated item tokenization, leading to the severe popularity bias. 

\begin{figure*}[t]
    \centering
    \includegraphics[width=\textwidth]{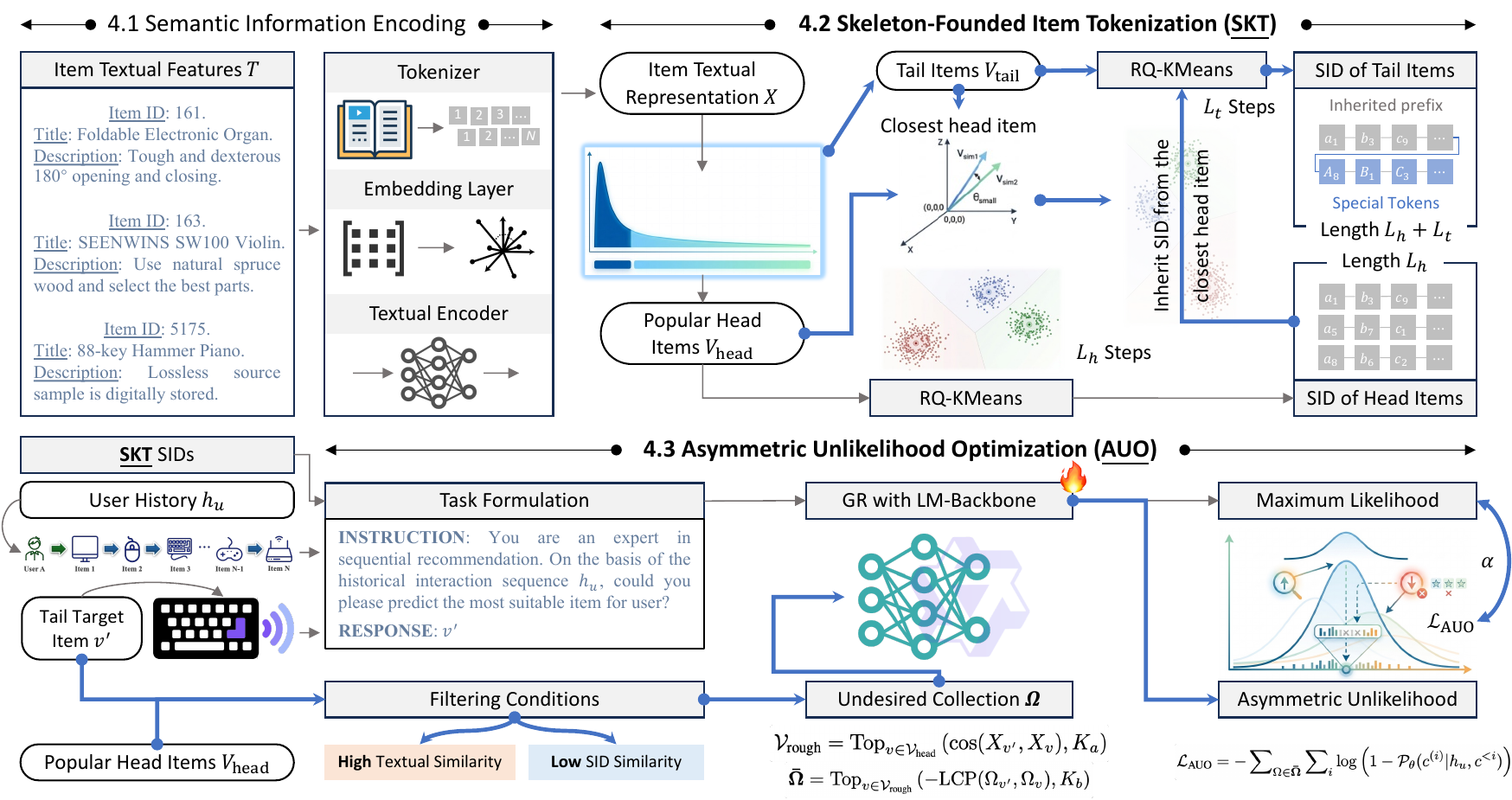}
    \caption{Overview of the Ghost model. 
    First, textual representations are encoded based on item features. After categorizing items into head and tail sets, \modela assigns SIDs via RQ-KMeans, allowing tail items to inherit prefixes from their closest head items. At last, \modelb optimizes the GR model by penalizing an undesired collection customized for each tail item, alongside standard MLE training.
    }
    \label{fig:method}
    \vspace{-0.2cm}
\end{figure*}

\section{Methodology}
With a focused effort to the two diagnosed fundamental factors, the \model model is developed by designing the skeleton-founded item tokenization (SKT) and the asymmetric unlikelihood optimization (AUO). Procedurally, the overview of \model is illustrated in Figure \ref{fig:method}. In particular, SKT subtly reduces unpredictable branching points by formulating a mechanism for SID inheritance from head to tail items. Hence, the bias amplification effect of undifferentiated tokenization is suppressed. Afterwards, based on the undesired collection of head item tokens, AUO is able to reallocate the supervision signals and thus rescue the gradient starvation issue of tail item tokens. 

\subsection{Skeleton-Founded Item Tokenization}
Current GRs \cite{tiger,lcrec,ed2} mostly adopt standard vector quantization techniques, such as RQ-VAE \cite{rqvae} and RQ-KMeans \cite{rqkmeans}, to allocate item SIDs. These standard approaches are agnostic to item popularity disparities, where head and tail items are indiscriminately processed. Therefore, the generated SIDs contain unstructured branching points where head and tail item tokens compete. To overcome this limitation, we propose the skeleton-founded item tokenization (SKT). In detail, SKT circumvents the chaotic competition between head and tail item tokens by specifying the position of the branching point. This approach makes it possible to capture the occurring patterns of tail items without having them drowned out by their popular counterparts. 

To be more specific, SKT assigns the SIDs for head and tail items asynchronously. Firstly, given the head item set $\mathcal V_{\text{head}}$ and the corresponding representations, SKT adopts the RQ-KMeans algorithm to generate the head item SIDs. For head item $v$ with representation $X_v$, the SID generation follows,
\begin{equation}
\label{eq:head_item_sid}
\begin{split}
    c_{v}^{(i)} &= \arg\min_{n\in\{1, 2,\dots,N\}} \| {r}_{v}^{(i)}-{\mu}_{n}^{(i)} \|_{2}^{2}, \\
    {r}_{v}^{(i+1)} &= {r}_{v}^{(i)} - {\mu}_{c_{v}^{(i)}}^{(i)}, \quad \text{for } i = 1, 2, \dots, L^{h}.
\end{split}
\end{equation}
$L^h$ is the SID length for head items. Initialized with $r_v^{(1)}=X_v$, the head item $v$ is assigned SID $\Omega_{v} = (c_{v}^{(1)},c_{v}^{(2)},\dots,c_{v}^{(L^h)})$. Afterwards, for the tail item $v'$ with representation $X_{v'}$, SKT begins by retrieving the head item with the highest semantic similarity, denoted as $v^*$. Subsequently, the tail item $v'$ inherits the first $L^h$ tokens from $\Omega_{v^*}$ and then additionally generates $L^t$ SID tokens. Similar to Eq.\eqref{eq:head_item_sid}, the additional SID generation for tail item $v'$ is represented below,
\begin{equation}
\label{eq:tail_item_sid}
\begin{split}
    c_{v'}^{(j)} &= \arg\min_{n\in\{1, 2,\dots,N\}} \| {r}_{v'}^{(j)}-{\mu}_{n}^{(j)}\|_{2}^{2}, \\
    {r}_{v'}^{(j+1)} &= {r}_{v'}^{(j)} - {\mu}_{c_{v'}^{(j)}}^{(j)}, \quad \text{for } j = 1, 2, \dots, L^{t}.
\end{split}
\end{equation}
Differing from the SID generation of head items, Eq.\eqref{eq:tail_item_sid} is initialized with $$r_{v'}^{(1)}=X_{v'}-\sum\nolimits_i\mu_{c_{v^*}^{(i)}}^{(i)},$$ and thus the tail item $v'$ is indexed with SID $\Omega_{v'}=(c_{v^*}^{(1)},\dots,c_{v^*}^{(L^h)},c_{v'}^{(1)},\cdots,c_{v'}^{(L^t)}).$ 

Conceptually, the SID collection of head items $\{\Omega_v|v\in\mathcal V_{\text{head}}\}$ serves as the skeleton of the whole SID system and essentially tessellates the SID space into several balanced partitions. On one hand, by inheriting the tail item SID from that of the closest head item, SKT maintains the correlation between items. Therefore, the basic principle of SID-based GRs, that similar items are prone to share similar SIDs, still holds. On the other hand, SKT unifies the branching point between head and tail item at step $(L^h+1)$, subtly reducing the bias amplification effect of undifferentiated item tokenization. Moreover, regarding the expressiveness of our SIDs, SKT captures the distinctiveness of the tail item $v'$ compared to its closest head item $v^*$ through $L^t$ additional SID tokens, which elegantly leverages the strong modeling capability on head items towards tail item improvement.

\subsection{Asymmetric Unlikelihood Optimization}
Based on the novel SIDs generated by SKT, \model further designs the asymmetric unlikelihood optimization (AUO) to rectify the gradient starvation issue of tail item tokens. As analyzed in Section \ref{sec:mle_opt}, the optimization of current GRs relies on MLE, where the tail items seldom serve as right answers. As a result, the optimized GR models prefer repeating head items, since they are never explicitly penalized for over-estimating statistically popular but contextually inappropriate head item tokens.

Inspired by the philosophy of unlikelihood training \cite{unlikelihood_0,unlikelihood_1}, AUO modulates the supervision signals by actively imposing penalties if the GR model generates notoriously undesirable tokens. Opposite to the standard NLL loss $\mathcal L_{\rm NLL}$, the AUO loss $\mathcal L_{\text{AUO}}$ can be concisely defined as follows,
\begin{equation}
    \mathcal L_{\text{AUO}}=-\sum\nolimits_{\Omega\in\mathbf{\bar\Omega}}\sum\nolimits_{i}\log \left(1-\mathcal P_{\theta}(c^{(i)}|h_{u},c^{<i})\right),\label{eq:auo}
\end{equation}
where $\bar{\mathbf\Omega}$ denotes the SID collection of undesired items. Notably, the AUO loss $\mathcal L_{\text{AUO}}$ is customized for tail items rather than the whole item set. For tail item $v'$ with representation $X_{v'}$ and SID $\Omega_{v'}$, \model first retrieves the top similar items from the popular head collection $\mathcal V_{\text{head}}$, which functions as a rough selection of undesired items. Then, \model further refines the rough candidates $\mathcal V_{\text{rough}}$ according to their SID distance from $\Omega_{v'}$, corresponding to a finer filtering of notoriously undesired items. Formally, the construction of $\bar{\mathbf\Omega}$ can be represented as follows,
\begin{equation}
\begin{split}
    \mathcal V_{\text{rough}} &= \mathop{\rm Top}_{v\in\mathcal V_{\text{head}}}\left(\cos(X_{v'},X_v),K_a\right), \\
    \mathbf{\bar\Omega} &= \mathop{\rm Top}_{v\in\mathcal V_{\text{rough}}}\left(-{\rm LCP}(\Omega_{v'},\Omega_{v}),K_b\right),
\end{split}
\end{equation}
where the ${\rm{Top}}(f(\cdot,\cdot),K)$ operator returns the $K$ objects with the highest $f$ function value, the $\rm LCP(\cdot,\cdot)$ operator returns the length of the \textit{Longest Common Prefix} between the two input sequences, and $K_a,K_b$ are two hyper-parameters controlling the candidate scale of the undesired item selection. 

For tail item $v'$, the undesired collection $\mathbf{\bar\Omega}_{v'}$ includes several head items whose representations are similar to $X_{v'}$, while their SIDs are divergent from $\Omega_{v'}$. Hence, AUO is able to inhibit the GR model from being hijacked by the popular head items that are similar to the ground-truth tail item in terms of semantic representation. In summary, the specialized asymmetric unlikelihood optimization is combined with MLE optimization, and the overall objective function of \model is defined as,
\begin{equation}
    \mathcal L_{\text{All}}=\mathcal L_{\text{NLL}} + \alpha\cdot\mathcal L_{\text{AUO}},\label{eq:overall}
\end{equation}
where $\alpha$ is the weighted parameter. Implementation details of Ghost are introduced in Appendix \ref{app:detail}.

\subsection{Theoretical Analysis}
Within SKT, since a tail item $v'$ inherits the prefix skeleton of its closest head item $v^*$, the branching point of item recommendation is uniformly deferred to step $(L^h+1)$. By establishing a singular, predictable branching point, SKT exponentially restricts the local amplification factor in LEMMA 2.

\textbf{LEMMA 3 (Mitigation of Bias Amplification).} \textit{The generation probability of the tail item $v'$ is insulated from the multi-step geometric suppression, and its deviation from the true data distribution $\mathcal{P}_d$ is governed exclusively by a localized factor,}
\begin{equation}
\label{eq:lemma3}
\begin{split}
    \mathcal{P}_{\theta}(v'|h_u) 
    &= \prod_{j=1}^{L^h+L^t} \mathcal{P}_{\theta}(c_{v'}^{(j)}|h_u, c^{<j}) \\
    &\approx (\gamma_{\textit{EOS}})^{-1} \cdot \prod_{j=1}^{L^h+L^t} \mathcal{P}_d(c_{v'}^{(j)}|h_u, c^{<j}).
\end{split}
\end{equation}
\textit{where $\gamma_{\text{EOS}} \geq 1$ represents the head-dominance factor confined to the $(L^h + 1)$-th generative recommendation step, in which the tokens of tail item $v'$ compete exclusively against the EOS token.}

Contrasting Eq.\eqref{eq:lemma3} with Eq.\eqref{eq:lemma2}, we can notice that by transforming a geometric suppression $\mathcal{O}(\gamma_{\text{min}}^{z})$ into a single-step discrepancy $\mathcal{O}(\gamma_{\textit{EOS}})$, SKT effectively halts the amplification of popularity bias during the generative recommendation process. Regarding the overall objective defined in Eq.\eqref{eq:overall}, a comprehensive gradient analysis is conducted to uncover its rationale. While MLE subjects tail tokens to gradient starvation through the suppressive $-\mathcal P_\theta(c^{(i)}_{\text{tail}}|h_u,c^{<i})\cdot X_{h_u}$ term, the Ghost model provides a powerful \textit{Rescue Force} on the basis of AUO. 

\textbf{LEMMA 4 (Gradient Rescue based on AUO).} 
\textit{For false positive head tokens $c^-_{\text{head}}\in\mathbf{\bar\Omega}$, the gradient correctly cancels the denominator via the softmax derivative for its own direct penalty. Crucially, it also systematically absorbs the cross-penalization dynamics originating from other tokens within the undesired set. Formally, the gradient is,}
\begin{equation}
\begin{split}
    \Delta e_{c^-_{\text{head}}} 
    &\propto -\underbrace{(1 + \alpha) \mathcal{P}_{\theta}(c^-_{\text{head}})\cdot X_{h_u}}_{\text{Targeted Repulsion}} \\
    &\quad + \underbrace{\alpha \sum_{c_j \in \bar{\mathbf{\Omega}} \setminus \{c^-_{\text{head}}\}} \frac{\mathcal{P}_{\theta}(c_j) \mathcal{P}_{\theta}(c^-_{\text{head}})}{1 - \mathcal{P}_{\theta}(c_j)}\cdot X_{h_u}}_{\text{Cross-Penalization Offset}}.
\end{split}
\end{equation}
\textit{For a token $c_{\text{tail}}$ of the target tail item, the expected update gradient is reformulated as,}
\begin{equation}
\begin{split}
    \Delta e_{c_{\text{tail}}} 
    &\propto \underbrace{-\mathcal{P}_{\theta}(c_{\text{tail}})\cdot X_{h_u}}_{\text{MLE Push}} \\
    &\quad + \underbrace{\alpha \sum_{c_j \in \bar{\mathbf{\Omega}}} \frac{\mathcal{P}_{\theta}(c_j)\mathcal{P}_{\theta}(c_{\text{tail}})}{1 - \mathcal{P}_{\theta}(c_j)}\cdot X_{h_u}}_{\text{AUO Rescue}}.
\end{split}
\end{equation}
By actively penalizing the undesired head tokens in $\mathbf{\bar{\Omega}}$, the Jacobian of the AUO loss systematically redistributes positive probability mass to the remaining tokens. This structural rescue ensures that tail tokens receive active, rational parameter updates, preventing them from being indefinitely pushed away from the user intent space as mere trivial negative samples. Detailed derivations of LEMMA 4 and LEMMA 5 are presented in Appendix \ref{app:theory} as well.

\section{Experiment}\label{sec:experiment}
$\bullet$ \textbf{Dataset}. This study evaluates \model 
and baselines on three public benchmarks extracted from the Amazon platform, i.e., \textit{"Musical \underline{Ins}truments"}, \textit{"\underline{Arts}, Crafts and Sewing"}, and \textit{"Video \underline{Games}"}.

$\bullet$ \textbf{Baseline}. The evaluation introduces three SOTA GRs (LETTER \cite{letter}, LC-Rec \cite{lcrec}, and ED$^2$ \cite{ed2}) as the standard baselines. IFairLRS \cite{ifair}, which pioneers the investigation of popularity bias in GRs, serves as a competitive baseline. In detail, \textit{RW} and \textit{RR} refer to the \textit{re-weighting only} and \textit{re-weight \& re-ranking} strategies. Moreover, to provide a comprehensive comparison, this study replaces each popular head item with the most similar tail item according to a pre-defined probability. The modified interaction sequence is either appended to the original training set or substituted for the original sequence, corresponding to the baselines denoted as \textit{Augmentation} and \textit{Substitution} (The details of baselines are presented in Appendix \ref{app:baseline}).

$\bullet$ \textbf{Metric}. Following well-established benchmarks \cite{lcrec,handbook}, this study adopts Hit-Rate (HR) and normalized discounted cumulative gain (NDCG) to evaluate the recommendation performance. Besides, this study adopts mean group unfairness (MGU) and average recommendation popularity (ARP) to quantify the fairness and the average popularity of the recommendation results \cite{2_8_split}. To reflect the proximity to Pareto optimality, this study defines the comprehensively normalized score (CNS) as follows, based on the Min-Max normalized values of HR, NDCG, MGU, and ARP metrics,
\begin{equation}
\begin{split}
    \text{CNS} &= \big( \overline{\text{HR}}_{\text{All}} + \overline{\text{HR}}_{\text{Tail}} + \overline{\text{NDCG}}_{\text{All}} \\
    &\quad + \overline{\text{NDCG}}_{\text{Tail}} + \overline{\text{MGU}} + \overline{\text{ARP}} \big) / 6.
\end{split}
\end{equation}
The detailed introduction of the datasets and metrics is presented in Appendix \ref{app:detail}.

\begin{table*}[t]
\centering
\setlength{\belowcaptionskip}{0.1cm}
\caption{
Performance comparison of Ghost and baselines across three datasets. 
The best, the runner-up, and the third-best results are highlighted in \colorbox[HTML]{B2DFDB}{\textbf{bold}}, \colorbox[HTML]{C5CAE9}{\underline{underlined}}, and \colorbox[HTML]{E6F2FF}{colored} fonts. 
}
\label{tab:main}
\renewcommand{\arraystretch}{1.15} 
\resizebox{\textwidth}{!}{
\begin{tabular}{c|c|cccccccccccc|cc}
\toprule
\multirow{2}{*}{\textbf{Dataset}} & \textbf{Metric} & \multicolumn{2}{c}{\textbf{HR@5}$~\uparrow$} & \multicolumn{2}{c}{\textbf{HR@10}$~\uparrow$} & \multicolumn{2}{c}{\textbf{NDCG@5}$~\uparrow$} & \multicolumn{2}{c}{\textbf{NDCG@10}$~\uparrow$} & \multirow{2}{*}{\textbf{MGU@5}$~\downarrow$} & \multirow{2}{*}{\textbf{MGU@10}$~\downarrow$} & \multirow{2}{*}{\textbf{ARP@5}$~\downarrow$} & \multirow{2}{*}{\textbf{ARP@10}$~\downarrow$} & \multirow{2}{*}{\textbf{CNS@5}$~\uparrow$} & \multirow{2}{*}{\textbf{CNS@10}$~\uparrow$} \\ \cmidrule(lr){3-4} \cmidrule(lr){5-6} \cmidrule(lr){7-8} \cmidrule(lr){9-10}
 & \textbf{Model} & \textbf{All} & \textbf{Tail} & \textbf{All} & \textbf{Tail} & \textbf{All} & \textbf{Tail} & \textbf{All} & \textbf{Tail} & & & & & & \\ 
\midrule
\midrule
\multirow{9}{*}{\textbf{Ins}} 
 & LETTER & 0.0593 & 0.0025 & 0.0662 & 0.0044 & 0.0530 & 0.0016 & 0.0553 & 0.0023 & 0.1965 & \third{0.1442} & \first{\textbf{181.2628}} & \first{\textbf{151.7149}} & 0.2457 & 0.2714 \\ 
 & LC-Rec & 0.0870 & 0.0027 & 0.1046 & 0.0040 & 0.0748 & 0.0021 & 0.0804 & 0.0025 & 0.3458 & 0.3380 & 365.4217 & 338.5304 & 0.3214 & 0.3134 \\ 
 & ED$^2$ & \first{\textbf{0.0898}} & 0.0067 & \second{\underline{0.1068}} & 0.0102 & \second{\underline{0.0765}} & 0.0043 & \second{\underline{0.0820}} & 0.0055 & 0.2551 & 0.2305 & 279.8572 & 237.9431 & 0.6067 & 0.6217 \\ 
 \cmidrule(lr){2-16}
 & Augmentation & \third{0.0875} & \third{0.0071} & 0.1039 & \third{0.0105} & \third{0.0759} & \third{0.0051} & \third{0.0812} & \third{0.0061} & 0.2383 & 0.2049 & 271.1657 & 242.7466 & \third{0.6327} & \second{\underline{0.6323}} \\ 
 & Substitution & 0.0780 & 0.0058 & 0.0914 & 0.0084 & 0.0645 & 0.0039 & 0.0689 & 0.0048 & \first{\textbf{0.0310}} & \first{\textbf{0.0114}} & \third{253.3299} & \third{229.5070} & 0.5714 & 0.5601 \\ 
 & IFairLRS-RW & \second{\underline{0.0888}} & 0.0058 & \first{\textbf{0.1072}} & \second{\underline{0.0081}} & \first{\textbf{0.0767}} & 0.0043 & \first{\textbf{0.0826}} & 0.0050 & 0.3322 & 0.3226 & 328.5940 & 299.5196 & 0.5007 & 0.4887 \\ 
 & IFairLRS-RR & 0.0853 & \second{\underline{0.0100}} & \third{0.1053} & \second{\underline{0.0110}} & 0.0736 & \first{\textbf{0.0078}} & 0.0800 & \second{\underline{0.0081}} & \third{0.1578} & 0.2330 & 301.2252 & 286.8023 & \second{\underline{0.7462}} & \third{0.6286} \\ 
 & Ghost (Ours) & 0.0864 & \first{\textbf{0.0117}} & 0.1017 & \first{\textbf{0.0173}} & 0.0755 & \first{\textbf{0.0078}} & 0.0805 & \first{\textbf{0.0097}} & \second{\underline{0.0596}} & \second{\underline{0.0958}} & \second{\underline{248.8179}} & \second{\underline{209.8877}} & \first{\textbf{0.8974}} & \first{\textbf{0.8694}} \\ 
\midrule
\midrule
\multirow{9}{*}{\textbf{Arts}} 
 & LETTER & 0.0448 & 0.0062 & 0.0521 & 0.0090 & 0.0378 & 0.0040 & 0.0401 & 0.0050 & 0.1768 & \third{0.1283} & \second{\underline{85.6235}} & \second{\underline{73.2413}} & 0.2108 & 0.2693 \\ 
 & LC-Rec & \first{\textbf{0.0885}} & 0.0188 & \first{\textbf{0.1095}} & 0.0269 & \first{\textbf{0.0737}} & 0.0133 & \first{\textbf{0.0805}} & 0.0159 & 0.2992 & 0.2905 & 135.2274 & 119.2569 & 0.5518 & 0.5548 \\ 
 & ED$^2$ & 0.0796 & 0.0105 & 0.0993 & 0.0172 & 0.0656 & 0.0073 & 0.0719& 0.0095 & 0.2960 & 0.2879 & 148.5051 & 129.2274 & 0.3331 & 0.3548 \\ 
 \cmidrule(lr){2-16}
 & Augmentation & \third{0.0839} & \second{\underline{0.0218}} & \third{0.1021} & \second{\underline{0.0306}} & {0.0697} & \second{\underline{0.0159}} & {0.0756} & \second{\underline{0.0187}} & \second{\underline{0.1572}} & \third{0.1299} & 112.3944 & 98.8012 & \second{\underline{0.6946}} & \third{0.7258} \\ 
 & Substitution & 0.0732 & \third{0.0215} & 0.0915 & \third{0.0305} & 0.0581 & \third{0.0155} & 0.0640 & \third{0.0184} & \first{\textbf{0.0490}} & \first{\textbf{0.0531}} & \third{96.9228} & \third{88.0055} & \third{0.6910} & \second{\underline{0.7269}} \\ 
 & IFairLRS-RW & \second{\underline{0.0845}} & 0.0138 & \second{\underline{0.1022}} & 0.0191 & \second{\underline{0.0706}} & 0.0102 & \second{\underline{0.0763}} & 0.0119 & 0.3076 & 0.2925 & 141.4042 & 122.9239 & 0.4359 & 0.4270 \\ 
 & IFairLRS-RR & 0.0827 & 0.0179 & 0.1005 & 0.0248 & 0.0696 & 0.0130 & 0.0753 & 0.0152 & 0.2255 & 0.2103 & 134.5895 & 117.1711 & 0.5447 & 0.5485 \\ 
 & Ghost (Ours) & 0.0831 & \first{\textbf{0.0296}} & 0.1000 & \first{\textbf{0.0393}} & \second{\underline{0.0706}} & \first{\textbf{0.0213}} & \third{0.0760} & \first{\textbf{0.0245}} & \third{0.1763} & 0.1621 & \first{\textbf{70.5222}} & \first{\textbf{65.1832}} & \first{\textbf{0.8828}} & \first{\textbf{0.8780}} \\ 
\midrule
\midrule
\multirow{9}{*}{\textbf{Games}} 
 & LETTER & 0.0264 & 0.0063 & 0.0375 & 0.0096 & 0.0187 & 0.0044 & 0.0222 & 0.0055 & 0.1727 & 0.1396 & \second{\underline{119.9021}} & \second{\underline{108.3559}} & 0.2458 & 0.2652 \\ 
 & LC-Rec & \first{\textbf{0.0636}} & 0.0148 & \first{\textbf{0.0938}} & 0.0244 & \first{\textbf{0.0438}} & 0.0091 & \first{\textbf{0.0536}} & 0.0122 & 0.2907 & 0.2740 & 164.9579 & 150.6237 & 0.5352 & 0.5499 \\ 
 & ED$^2$ & 0.0572 & 0.0086 & 0.0854 & 0.0150 & 0.0393 & 0.0052 & \third{0.0484} & 0.0072 & 0.3110 & 0.2925 & 188.5855 & 170.8846 & 0.3052 & 0.3284 \\ 
 \cmidrule(lr){2-16}
 & Augmentation & 0.0559 & 0.0173 & 0.0822 & 0.0282 & 0.0384 & 0.0113 & 0.0469 & 0.0148 & 0.1526 & \third{0.1210} & 138.5779 & 125.0857 & 0.6701 & 0.6978 \\ 
 & Substitution & 0.0448 & \third{0.0199} & 0.0683 & \third{0.0311} & 0.0309 & \third{0.0131} & 0.0384 & \third{0.0167} & \second{\underline{0.0783}} & \second{\underline{0.0824}} & \third{121.4722} & \third{111.5519} & \third{0.7074} & \third{0.7093} \\ 
 & IFairLRS-RW & \second{\underline{0.0605}} & 0.0134 & \second{\underline{0.0907}} & 0.0228 & \second{\underline{0.0416}} & 0.0087 & 0.0513 & 0.0117 & 0.2869 & 0.2710 & 166.8988 & 153.0392 & 0.4878 & 0.5103 \\ 
 & IFairLRS-RR & \third{0.0581} & \second{\underline{0.0229}} & \third{0.0877} & \second{\underline{0.0315}} & \third{0.0402} & \second{\underline{0.0153}} & \second{\underline{0.0497}} & \second{\underline{0.0181}} & \third{0.1051} & 0.1449 & 148.2428 & 141.5092 & \second{\underline{0.8073}} & \second{\underline{0.7229}} \\ 
 & Ghost (Ours) & 0.0562 & \first{\textbf{0.0257}} & 0.0840 & \first{\textbf{0.0392}} & 0.0390 & \first{\textbf{0.0167}} & 0.0479 & \first{\textbf{0.0218}} & \first{\textbf{0.0743}} & \first{\textbf{0.0484}} & \first{\textbf{111.4375}} & \first{\textbf{106.0538}} & \first{\textbf{0.9349}} & \first{\textbf{0.9406}} \\ 
\bottomrule
\end{tabular}
}
\vspace{-0.4cm}
\end{table*}
\subsection{Main Result}
To demonstrate the effectiveness of \model in mitigating popularity bias, we compare its performance against SOTA baselines across three public datasets. The results are summarized in Table \ref{tab:main}. 

In general, \model can effectively approach the Pareto frontier while jointly considering overall recommendation performance, tail recommendation performance, and recommendation fairness. In detail, we can draw the following three observations. \textit{\textbf{First}}, \model is primarily characterized by its significant performance gains in tail-item recommendation. Across three datasets, Ghost outperforms standard GRs (i.e., LETTER, LC-Rec, and ED$^2$) by increasing tail HR and NDCG by 63.91\% and 70.66\% on average. Compared to existing popularity debiasing methods, it delivers an average increase of 28.39\% and 15.04\% for tail HR and NDCG, with maximum improvements up to 57.24\%. \textit{\textbf{Second}}, \model effectively suppresses the over-recommendation of head items. It lowers the MGU by an average of 55.76\% compared to standard GRs, and outperforms the most competitive popularity debiasing baseline, IFairLRS-RR, by further reducing the MGU by 16.68\% on average (up to a maximum of 66.60\%). \textit{\textbf{Third}}, with respect to overall recommendation performance, \model exhibits an acceptable degradation compared to standard models. Relative to the strongest baseline, LC-Rec, the overall HR and NDCG of \model drop by only 7.46\% and 6.34\% on average. For context, the performance drops among all popularity debiasing methods in Table \ref{tab:main} range between 2.46\% and 18.90\% for HR, and between 2.24\% and 21.28\% for NDCG. Taken together, Ghost yields an average CNS gain of 16.81\%, peaking at 22.06\%, which indicates its capability to substantially approach the Pareto optimal state for the GRs popularity bias issue.

\begin{table*}[t]
\centering
\setlength{\belowcaptionskip}{0.1cm}
\caption{Performance comparison of ablation study on \textit{Ins} dataset.
\textit{RQK-4/6} denotes a LC-Rec model adopted 4/6-tokens SIDs provided by RQ-KMeans, respectively. \textit{RQK-4-6} adopts 4-tokens SIDs for head items while 6-tokens SIDs for tail items, while without the inheriting process in SKT.
}
\label{tab:ablation}
\renewcommand{\arraystretch}{1.1} 
\resizebox{\textwidth}{!}{
\begin{tabular}{c|cccccccccccc}
\toprule
\multirow{2}{*}{\textbf{Model}} & \multicolumn{2}{c}{\textbf{HR@5$~\uparrow$}} & \multicolumn{2}{c}{\textbf{HR@10$~\uparrow$}} & \multicolumn{2}{c}{\textbf{NDCG@5$~\uparrow$}} & \multicolumn{2}{c}{\textbf{NDCG@10$~\uparrow$}} & \multirow{2}{*}{\textbf{MGU@5$~\downarrow$}} & \multirow{2}{*}{\textbf{MGU@10$~\downarrow$}} & \multirow{2}{*}{\textbf{ARP@5$~\downarrow$}} & \multirow{2}{*}{\textbf{ARP@10$~\downarrow$}} \\ 
\cmidrule(lr){2-3} \cmidrule(lr){4-5} \cmidrule(lr){6-7} \cmidrule(lr){8-9}
 & \textbf{All} & \textbf{Tail} & \textbf{All} & \textbf{Tail} & \textbf{All} & \textbf{Tail} & \textbf{All} & \textbf{Tail} & & & & \\ 
\midrule
Ghost & 0.0864 & \first{\textbf{0.0117}} & 0.1017 & \first{\textbf{0.0173}} & 0.0755 & \first{\textbf{0.0078}} & 0.0805 & \first{\textbf{0.0097}} & \second{\underline{0.0596}} & \second{\underline{0.0958}} & \first{\textbf{248.8179}} & \first{\textbf{209.8877}} \\ 
w/o AUO & 0.0849 & \second{\underline{0.0111}} & 0.1003 & \second{\underline{0.0161}} & 0.0733 & \second{\underline{0.0075}} & 0.0782 & \second{\underline{0.0091}} & \first{\textbf{0.0035}} & \first{\textbf{0.0112}} & \second{\underline{270.4236}} & \second{\underline{235.9759}} \\ 
w/o SKT & \third{0.0920} & 0.0060 & \third{0.1102} & \third{0.0095} & \first{\textbf{0.0798}} & 0.0047 & \second{\underline{0.0856}} & \third{0.0058} & 0.3211 & 0.3056 & 387.0114 & 357.7677 \\ 
RQK-4 & \second{\underline{0.0929}} & 0.0059 & \second{\underline{0.1116}} & \third{0.0095} & \third{0.0796} & 0.0042 & \second{\underline{0.0856}} & 0.0054 & 0.3037 & 0.2885 & 304.3177 & \third{267.9522} \\ 
RQK-6 & \first{\textbf{0.0932}} & 0.0052 & \first{\textbf{0.1133}} & 0.0088 & \second{\underline{0.0797}} & 0.0038 & \first{\textbf{0.0862}} & 0.0049 & 0.3073 & 0.2902 & \third{303.1651} & 269.7153 \\ 
RQK-4-6 & 0.0883 & \third{0.0070} & 0.1080 & 0.0094 & 0.0762 & \third{0.0049} & 0.0826 & 0.0057 & \third{0.2268} & \third{0.2468} & 320.2727 & 288.6501 \\ 
\bottomrule
\end{tabular}
}
\vspace{-0.4cm}
\end{table*}
\subsection{Ablation Study}
To investigate the contributions of the core components in Ghost, an ablation study is conducted by removing specific modules. Based on Table \ref{tab:ablation}, the following insights can be drawn.
\textbf{\textit{First}}, bias amplification stemming from undifferentiated tokenization constitutes the critical cause of the GRs susceptibility to popularity bias. A performance comparison between Ghost and the \textit{w/o SKT} variant reveals that the inability to inhibit the unpredictable competition between head and tail tokens leads to a substantial decline in tail recommendation performance. The results from \textit{RQK-4}, \textit{RQK-6}, and \textit{RQK-4-6} serve as additional evidence supporting this finding. \textbf{\textit{Second}}, the supervisory signals corrected by AUO must be absorbed by the corresponding learnable parameters. By comparing the performance of Ghost, \textit{w/o AUO}, and \textit{w/o SKT}, one can notice that the model ability to mitigate popularity bias is limited when relying solely on AUO. However, once SKT introduces additional tokens for tail items, the corrective effect provided by AUO drives the model closer to Pareto optimality.

\begin{figure*}[h]
    \centering
    \includegraphics[width=\textwidth]{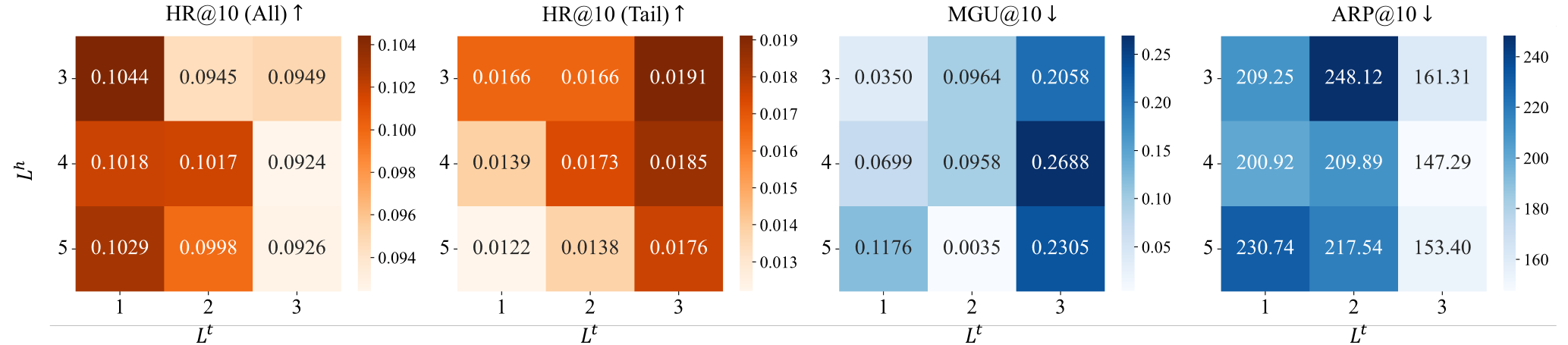}
    \caption{
    Analysis of SID lengths, including head length $L^h$ and additional length $L^t$ for tail items.
    }
    \label{fig:sid_length}
    \vspace{-0.2cm}
\end{figure*}

\subsection{SID Length Analysis}
As shown by Figure \ref{fig:sid_length}, we vary the skeleton length $L^h\in$\{3, 4, 5\} and the additional tail-specific length $L^t\in$\{1, 2, 3\}. The following insights can be drawn. A larger $L^t$ effectively mitigates popularity bias, as evidenced by the improvement in tail HR@10 and the reduction in ARP@10. However, this exploration comes at the expense of overall performance and fairness. This reveals a clear trade-off in identifier length allocation. Assigning longer sequences to tail items enhances their representational capacity and retrieval probability, but it simultaneously introduces noise that dilutes the learning of head items, thereby compromising overall performance. Furthermore, the skeleton length $L^h$ dictates the extent to which tail items inherit semantic prefixes from head items. Forcing tail items to strictly mirror head items over too many generative steps suppresses their unique semantic identities. 

\begin{table*}[htb]
\centering
\setlength{\belowcaptionskip}{0.1cm}
\caption{Performance comparison across different backbone scales. The best and the runner-up are highlighted in \colorbox[HTML]{B2DFDB}{\textbf{bold}}, \colorbox[HTML]{C5CAE9}{\underline{underlined}} fonts. }
\label{tab:scaling}
\renewcommand{\arraystretch}{1.15}
\resizebox{\textwidth}{!}{
\begin{tabular}{cccccccccccccc}
\toprule
\multirow{2}{*}{\textbf{{Scale}}} & \multirow{2}{*}{\textbf{Model}} & \multicolumn{2}{c}{\textbf{HR@5$~\uparrow$}} & \multicolumn{2}{c}{\textbf{HR@10$~\uparrow$}} & \multicolumn{2}{c}{\textbf{NDCG@5$~\uparrow$}} & \multicolumn{2}{c}{\textbf{NDCG@10$~\uparrow$}} & \multirow{2}{*}{\textbf{MGU@5$~\downarrow$}} & \multirow{2}{*}{\textbf{MGU@10$~\downarrow$}} & \multirow{2}{*}{\textbf{ARP@5$~\downarrow$}} & \multirow{2}{*}{\textbf{ARP@10$~\downarrow$}} \\ 
\cmidrule(lr){3-4} \cmidrule(lr){5-6} \cmidrule(lr){7-8} \cmidrule(lr){9-10}
 & & \textbf{All} & \textbf{Tail} & \textbf{All} & \textbf{Tail} & \textbf{All} & \textbf{Tail} & \textbf{All} & \textbf{Tail} & & & & \\ 
\midrule
\multirow{4}{*}{0.6B} 
 & LC-Rec & 0.0794 & 0.0015 & 0.0983 & 0.0021 & 0.0671 & 0.0010 & 0.0731 & 0.0011 & 0.3568 & 0.3550 & 433.3062 & 404.2913 \\ 
 & IFairLRS-RW & \second{\underline{0.0847}} & 0.0025 & \first{\textbf{0.1030}} & 0.0036 & \second{\underline{0.0725}} & 0.0018 & \second{\underline{0.0784}} & 0.0021 & 0.3483  & 0.3431 & 393.1610 & 360.2472 \\ 
 & IFairLRS-RR & 0.0819 & \second{\underline{0.0053}} & 0.1010 & \second{\underline{0.0060}} & 0.0705 & \second{\underline{0.0045}} & 0.0766 & \second{\underline{0.0047}} & \second{\underline{0.2414}} & \second{\underline{0.2918}} & \second{\underline{378.3699}} & \second{\underline{353.8215}} \\ 
 & Ghost & \first{\textbf{0.0848}} & \first{\textbf{0.0101}} & \second{\underline{0.1025}} & \first{\textbf{0.0148}} & \first{\textbf{0.0748}} & \first{\textbf{0.0069}} & \first{\textbf{0.0805}} & \first{\textbf{0.0084}} & \first{\textbf{0.0136}} & \first{\textbf{0.0208}} & \first{\textbf{239.9323}} & \first{\textbf{209.8331}} \\ 
\midrule
\multirow{4}{*}{1.7B} 
 & LC-Rec & 0.0855 & 0.0027 & 0.1025 & 0.0033 & 0.0731 & 0.0022 & 0.0786 & 0.0024 & 0.3531 & 0.3512 & 418.5129 & 386.3999 \\ 
 & IFairLRS-RW & \first{\textbf{0.0879}} & 0.0038 & \first{\textbf{0.1070}} & 0.0057 & \first{\textbf{0.0764}} & 0.0026 & \first{\textbf{0.0825}} & 0.0032 & 0.3459 & 0.3422 & 373.8739 & 341.3491 \\ 
 & IFairLRS-RR & 0.0847 & \second{\underline{0.0084}} & 0.1047 & \second{\underline{0.0090}} & 0.0734 & \second{\underline{0.0070}} & 0.0798 & \second{\underline{0.0071}} & \second{\underline{0.2052}} & \second{\underline{0.2676}} & \second{\underline{346.5825}} & \second{\underline{326.8923}} \\ 
 & Ghost & \second{\underline{0.0870}} & \first{\textbf{0.0110}} & \second{\underline{0.1059}} & \first{\textbf{0.0174}} & \second{\underline{0.0761}} & \first{\textbf{0.0081}} & \second{\underline{0.0821}} & \first{\textbf{0.0102}} & \first{\textbf{0.0262}} & \first{\textbf{0.0508}} & \first{\textbf{248.1768}} & \first{\textbf{224.2784}} \\ 
\midrule
\multirow{4}{*}{4B} 
 & LC-Rec & \first{\textbf{0.0937}} & 0.0057 & \second{\underline{0.1112}} & 0.0074 & \first{\textbf{0.0811}} & 0.0047 & \first{\textbf{0.0867}} & 0.0053 & 0.3363 & 0.3301 & 330.5470 & 301.0089 \\ 
 & IFairLRS-RW & \first{\textbf{0.0937}} & 0.0067 & \first{\textbf{0.1125}} & 0.0097 & \second{\underline{0.0800}} & 0.0051 & \second{\underline{0.0860}} & 0.0061 & 0.3160 & 0.3033 & 293.3685 & 264.1469 \\ 
 & IFairLRS-RR & 0.0884 & \second{\underline{0.0124}} & 0.1097 & \second{\underline{0.0148}} & 0.0764 & \second{\underline{0.0097}} & 0.0832 & \second{\underline{0.0105}} & \second{\underline{0.0871}} & \second{\underline{0.1836}} & \second{\underline{256.1628}} & \second{\underline{247.6822}} \\ 
 & Ghost & \second{\underline{0.0898}} & \first{\textbf{0.0136}} & 0.1074 & \first{\textbf{0.0190}} & 0.0782 & \first{\textbf{0.0101}} & 0.0838 & \first{\textbf{0.0118}} & \first{\textbf{0.0356}} & \first{\textbf{0.0211}} & \first{\textbf{239.5142}} & \first{\textbf{215.0566}} \\ 
\midrule
\multirow{4}{*}{8B} 
 & LC-Rec & \first{\textbf{0.0954}} & 0.0062 & \first{\textbf{0.1170}} & 0.0092 & \first{\textbf{0.0817}} & 0.0047 & \first{\textbf{0.0886}} & 0.0056 & 0.3233 & 0.3142 & 300.1242 & \second{\underline{268.9937}} \\ 
 & IFairLRS-RW & 0.0900 & 0.0066 & 0.1102 & 0.0088 & 0.0785 & 0.0048 & 0.0850 & 0.0055 & 0.3251 & 0.3177 & 317.3005 & 285.7003 \\ 
 & IFairLRS-RR & 0.0872 & \second{\underline{0.0120}} & 0.1076 & \second{\underline{0.0140}} & 0.0764 & \second{\underline{0.0095}} & 0.0830 & \second{\underline{0.0101}} & \second{\underline{0.1445}} & \second{\underline{0.2212}} & \second{\underline{289.5438}} & 273.9240 \\ 
 & Ghost & \second{\underline{0.0907}} & \first{\textbf{0.0150}} & \second{\underline{0.1109}} & \first{\textbf{0.0228}} & \second{\underline{0.0787}} & \first{\textbf{0.0106}} & \second{\underline{0.0851}} & \first{\textbf{0.0131}} & \first{\textbf{0.0479}} & \first{\textbf{0.0410}} & \first{\textbf{207.7282}} & \first{\textbf{181.4011}} \\ 
\bottomrule
\end{tabular}
}
\end{table*}
\subsection{Scaling Pattern Analysis}\label{app:scaling}
To further investigate the scalability of our proposed framework, we compare the performance of Ghost against various baselines across different LLM backbone scales (ranging from 0.6B to 8B). The results are summarized in Table \ref{tab:scaling}.

In general, Ghost consistently demonstrates superior popularity bias mitigation and long-tail item excavation capabilities regardless of the underlying backbone size. In detail, we can draw the following three observations. \textbf{\textit{First}}, Ghost is primarily characterized by its robust debiasing and tail-item recommendation performance across all scales. Compared to both the standard baseline LC-Rec and the fairness-aware baselines IFairLRS, Ghost consistently achieves the highest Tail HR and Tail NDCG, alongside the lowest ARP and MGU scores. For instance, at the 8B scale, Ghost substantially increases Tail HR@10 to 0.0228 compared to 0.0092 of LC-Rec, while drastically reducing ARP@10 to 181.4011 compared to 268.9937 of LC-Rec. \textbf{\textit{Second}}, with respect to overall recommendation performance, Ghost maintains highly competitive utility. Despite aggressively promoting tail items, the overall HR and NDCG metrics of Ghost remain closely comparable to those of the standard generative recommendation models and generally surpass the IFairLRS baselines. This indicates that Ghost achieves an excellent trade-off, mitigating popularity bias without inducing unacceptable degradation in general recommendation accuracy. \textbf{\textit{Third}}, Ghost exhibits strong scalability with respect to LLM backbone capacity. As the parameter size increases from 0.6B to 8B, Ghost effectively leverages the enhanced representation and reasoning capabilities of larger models to further boost its performance. Specifically, Ghost's Tail HR@10 steadily increases from 0.0148 at 0.6B to 0.0228 at 8B, while its ARP@10 progressively drops from 209.8331 to 181.4011. This demonstrates that larger backbones consistently empower the framework to deliver increasingly diverse, balanced, and fair recommendations.

\section{Head-to-Tail Ratio Analysis}\label{app:item_number}
\begin{figure*}[t]
    \centering
    \includegraphics[width=\textwidth]{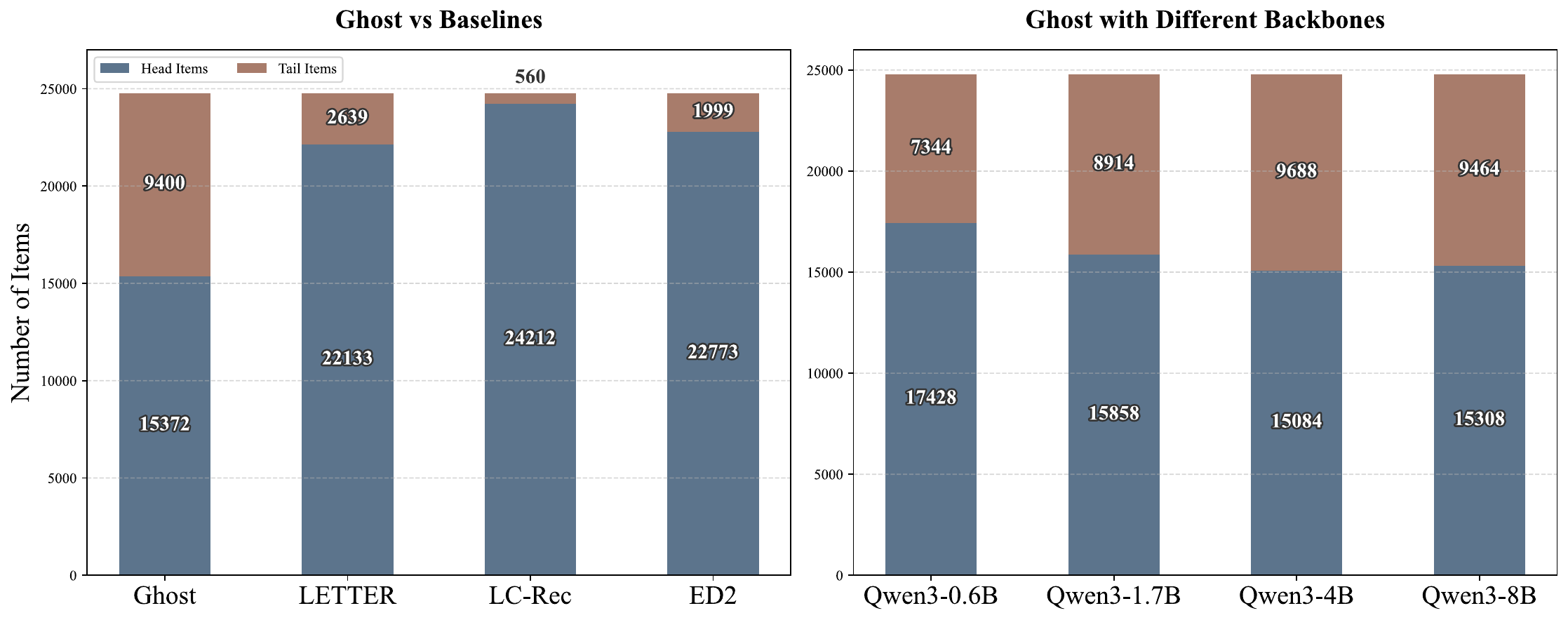}
    \caption{
    Numbers of head and tail items in the recommendation results provided by \textit{(left)} Ghost and baseline models, and \textit{(right)} Ghost with different backbones.
    }
    \label{fig:item_number}
    \vspace{-0.2cm}
\end{figure*}

In this section, we present the ratio of head items to tail items in the recommendation results provided by each GR model. Figure \ref{fig:item_number} illustrates the exposure distribution of head and tail items retrieved by Ghost and various baselines. While existing generative baselines like LETTER, LC-Rec, and ED2 exhibit severe popularity bias by disproportionately favoring head items. For example, LC-Rec retrieves 24,212 head items versus only 560 tail items. Ghost significantly alleviates this discrepancy, achieving a balanced distribution of 15,372 head and 9,400 tail items. Furthermore, evaluating Ghost across different Qwen3 backbone scales reveals a positive correlation between LLM capacity and long-tail item excavation. Upgrading the backbone from 0.6B to 4B parameters steadily increases the retrieval of tail items from 7,344 to 9,688, demonstrating that the enhanced representation capabilities of larger models effectively empower the framework to deliver diverse, balanced recommendations without overfitting to mainstream trends.

\section{Hyper-parameter Analysis}\label{app:hyper}

\begin{figure*}[t]
    \centering
    \includegraphics[width=\textwidth]{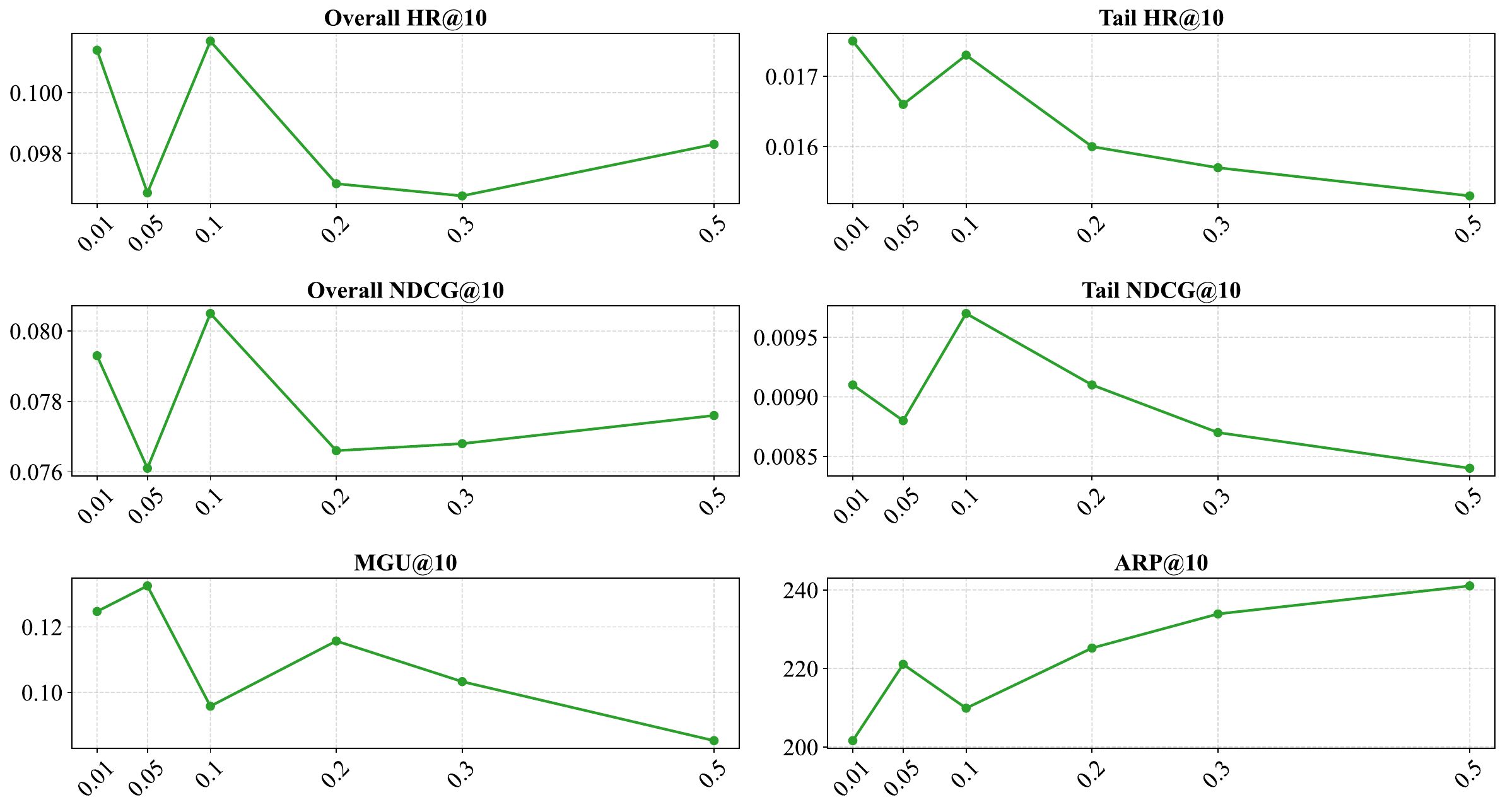}
    \caption{
    Tendency of \model performance on Ins dataset, under different AUO weights $\alpha$. The $x$-axis denotes the values of $\alpha$, and the $y$-axis is the metric values.
    }
    \label{fig:alpha}
    \vspace{-0.2cm}
\end{figure*}

\subsection{AUO Weight $\alpha$}
Here, we investigate the impact of the weight parameter $\alpha$ of AUO. Figure \ref{fig:alpha} investigates the sensitivity of the Ghost model to the AUO weight $\alpha$ on the Ins dataset. The results indicate that $\alpha$ plays a critical role in balancing recommendation accuracy and popularity bias mitigation. Specifically, $\alpha=0.1$ emerges as the optimal setting, achieving the highest Overall NDCG@10 and Tail NDCG@10 while maintaining highly competitive Overall and Tail HR@10. Crucially, at this optimal point, the ARP@10 experiences a local drop, confirming the model's capability to effectively surface less popular items without sacrificing accuracy. Conversely, increasing the weight beyond 0.1 (e.g., $0.2$ to $0.5$) consistently degrades both overall and tail-specific accuracy metrics, significantly diminishes MGU@10, and steadily increases ARP@10. This clear trend demonstrates that an excessively large AUO weight forces the model to over-prioritize mainstream items, thereby exacerbating popularity bias and compromising both long-tail excavation and overall recommendation quality.

\begin{figure*}[t]
    \centering
    \includegraphics[width=\textwidth]{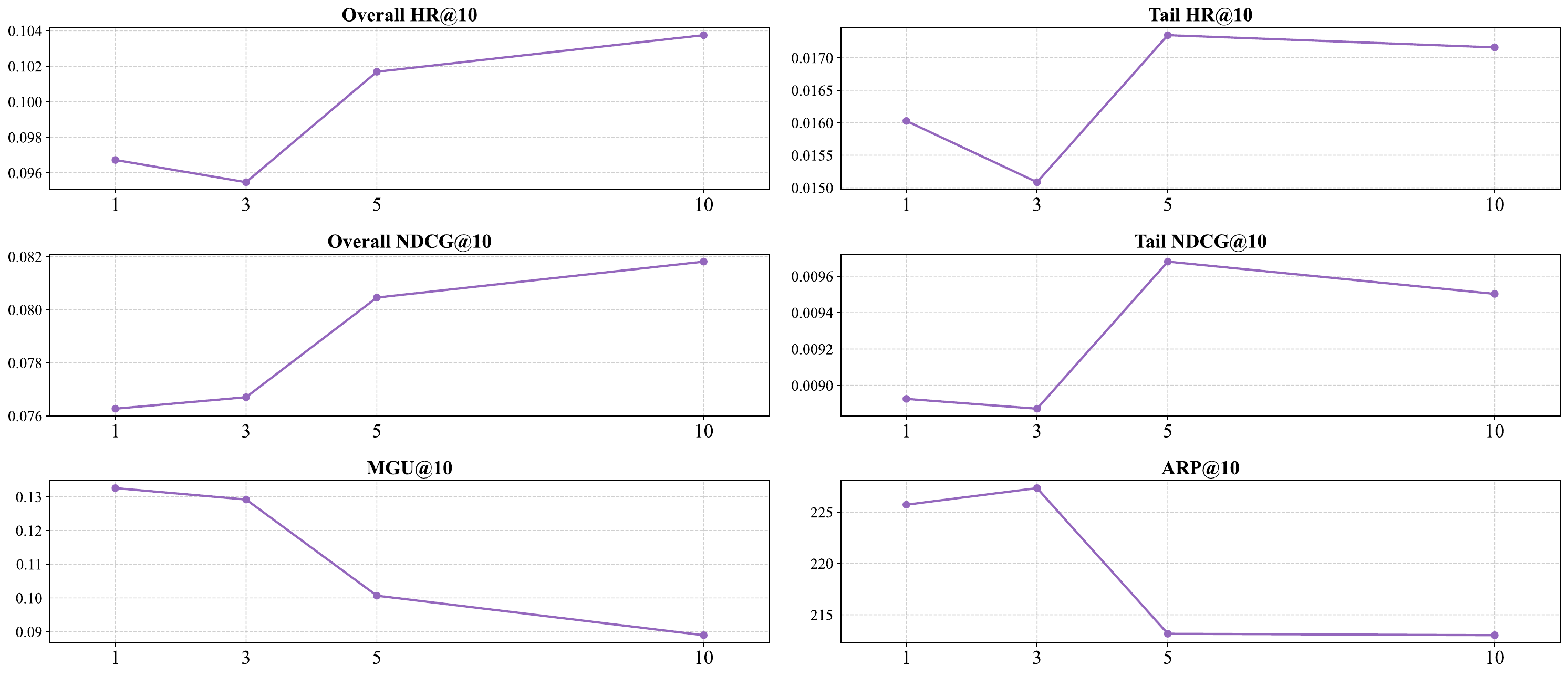}
    \caption{
    Tendency of \model performance on Ins dataset, under different undesired collection sizes. The $x$-axis denotes the size of undesired collection, and the $y$-axis is the metric values.
    }
    \label{fig:nega_size}
    \vspace{-0.2cm}
\end{figure*}

\subsection{Size of Undesired Collection $\mathbf{\bar{\Omega}}$}
Here, we investigate the impact of the undesired collection size. Figure \ref{fig:nega_size} investigates the impact of the undesired collection size $|\mathbf{\bar{\Omega}}|$ on the performance of the Ghost model using the Ins dataset. The empirical results demonstrate that the magnitude of $|\mathbf{\bar{\Omega}}|$ critically influences the trade-off between recommendation accuracy and popularity bias mitigation. Specifically, setting the collection size to 5 emerges as the optimal configuration; at this point, the model achieves peak effectiveness in excavating long-tail items, evidenced by the highest Tail HR@10 and Tail NDCG@10, alongside a sharp and favorable decline in ARP@10. Concurrently, overall accuracy metrics (Overall HR@10 and NDCG@10) experience substantial gains compared to smaller sizes. However, expanding the undesired collection further to 10 yields diminishing returns: while overall accuracy marginally increases, tail-specific metrics slightly degrade, and user coverage (MGU@10) continues a steady, negative decline. This indicates that a moderate undesired collection size effectively provides sufficient contrast against over-recommended popular items to surface relevant tail items, whereas an excessively large collection may introduce noise that compromises tail item retrieval and overall recommendation diversity across different users.

\section{Conclusion}
In this paper, we theoretically demonstrate that standard MLE training and undifferentiated item tokenization of current GRs inherently cause a token-level optimization flaw and multi-step bias amplification. Accordingly, we propose Ghost equipped with asymmetric unlikelihood optimization (AUO) and skeleton-founded tokenization (SKT). AUO provides explicit negative supervision to rescue tail tokens from gradient starvation, while SKT establishes unified branching points to halt item-level bias amplification. Extensive empirical evaluations confirm that Ghost effectively breaks the filter bubble and substantially promotes fairer long-tail recommendations with slight losses to overall recommendation utility, approaching Pareto optimality in GRs popularity debiasing.

\section*{Impact Statement}\label{app:impact}
This work aims to enhance the fairness and diversity of Generative Recommender Systems. By mitigating the fundamental causes of popularity bias, our proposed model effectively breaks the pervasive filter bubble. Ethically, this approach addresses the algorithmic "Matthew Effect," which typically over-represents trending items while severely marginalizing niche content. By promoting fairer long-tail recommendations without significantly compromising overall utility, this research fosters the responsible deployment of LLM-based recommendation technologies.

\clearpage
\bibliographystyle{IEEEtran}
\balance
\bibliography{ref}











\clearpage
\appendices
\onecolumn

\section{Notation}\label{app:notation}
The notations and corresponding descriptions are summarized in Table \ref{tab:notation}.
\begin{table}[htbp]
\centering
\caption{Summary of Mathematical and Model Notations}
\label{tab:notation}
\renewcommand{\arraystretch}{1.15}
\resizebox{\textwidth}{!}{
\begin{tabular}{cp{4.2cm}|cp{4.2cm}|cp{4.2cm}}
\toprule
\textbf{Symbol} & \textbf{Description} & \textbf{Symbol} & \textbf{Description} & \textbf{Symbol} & \textbf{Description} \\
\midrule
$K, J$ & Number of items and users in the system, respectively. & $L$ & Length of SIDs. & $\mathcal{V}_{\text{head}}$ & Set of popular head items. \\
$v_{k}, u_{j}$ & The $k$-th item and $j$-th user. & $\theta$ & Parameters of the Generative Recommender (GR) model. & $L^{h}$ & SID length assigned for head items (the skeleton length). \\
$h_{u_{j}}$ or $h_{u}$ & Historical behavior / item sequence of user $u_{j}$. & $\mathcal{P}_{\theta}$ & Token generation probability parameterized by $\theta$. & $v^{\prime}$ & A target tail item. \\
$l$ & Length of the item sequence. & $\mathcal{P}_{d}$ & True data distribution. & $v^{*}$ & The closest head item to tail item $v^{\prime}$ based on highest semantic similarity. \\
$T_{k}$ & Textual features (e.g., title, description) attached to item $v_{k}$. & $\mathcal{L}_{\text{NLL}}$ & Negative log-likelihood loss used for Maximum Likelihood Estimation (MLE). & $L^{t}$ & Additional SID tokens length specifically for tail items. \\
$f_{e}$ & Pre-trained textual encoder. & $\mathcal{D}$ & Training distribution of user-item interaction pairs. & $\gamma_{EOS}$ & Head-dominance factor at the $(L^{h}+1)$-th generative step against the EOS token. \\
$X_{v}, X_{k}$ & Semantic textual representation of item $v$ or $k$. & $X_{h_{u}}$ & Encoded representation of the user's historical behavior $h_{u}$. & $\overline{\mathbf{\Omega}}$ & SID collection of roughly selected undesired head items. \\
$\mu_{n}^{(i)}$ & The $n$-th embedding in the $i$-th codebook within the RQ-VAE. & $e_{c}$ & Embedding of token $c$. & $\mathcal{V}_{\text{rough}}$ & Rough candidate set of popular head items semantically similar to the tail item. \\
$r_{v}^{(i)}$ & Residual embedding at step $i$ during tokenization. & $c_{\text{head}}^{(i)}, c_{\text{tail}}^{(i)}$ & Candidate head and tail tokens competing at the $i$-th step. & $K_{a}, K_{b}$ & Hyper-parameters controlling the candidate scale for undesired item selection. \\
$c_{v}^{(i)}$ & The $i$-th SID token for item $v$. & $\gamma_{i}$ & Amplification factor at step $i$. & $\mathcal{L}_{\text{AUO}}$ & Asymmetric unlikelihood optimization (AUO) loss. \\
$\Omega_{v}$ & Complete SID of item $v$, indexed as $(c_{v}^{(1)}, c_{v}^{(2)}, \dots, c_{v}^{(L)})$. & $\mathcal{Z}$ & Set of steps during tail item generation where it competes against head tokens. & $\mathcal{L}_{\text{All}}$ & Overall optimization objective function of the Ghost model. \\
$c_{v}^{<i}$ & Sub-sequence of SID $\Omega_{v}$ before the $i$-th position. & $\gamma_{\text{min}}$ & Minimum factor across the competing steps in $\mathcal{Z}$. & $\alpha$ & Weighted parameter controlling the AUO loss. \\
\bottomrule
\end{tabular}
}
\end{table}

\section{Related Work}\label{app:related}
\subsection{Recommender Systems: Discriminative and Generative Paradigms}
Traditional recommender systems typically formulate recommendation as a discriminative task, relying heavily on discrete item IDs to capture user preferences by analyzing historical interactions \cite{ncf, sasrec, lightgcn, alibaba, youtube}. At the core of this pipeline, both users and items are projected into a shared low-dimensional latent space to learn dense embeddings, allowing the system to determine the final ranking by computing their pairwise similarities or interaction scores. To efficiently handle massive item catalogs, these systems typically employ a multi-stage pipeline, primarily consisting of candidate generation and ranking. Specifically, the candidate generation stage rapidly retrieves a coarse-grained subset of relevant items from the entire corpus, which are subsequently evaluated and sorted by a more complex ranking model to produce the final recommendations. 

Recently, fueled by the adoption of large language models (LLMs) \cite{GPT, llama, deepseek} as the underlying backbone, Generative Recommenders (GRs) have emerged as a transformative paradigm \cite{tiger, handbook}. Instead of modeling interaction probabilities directly, GRs replace traditional item IDs with Semantic IDs (SIDs), reframing recommendation as a unified end-to-end sequential generation process \cite{letter, lcrec, ed2, mql4grec}. To construct these SIDs, standard vector quantization techniques, such as VQ-VAE \cite{vqvae}, RQ-VAE \cite{rqvae}, and RQ-KMeans \cite{rqkmeans}, are widely utilized to convert continuous item embeddings into discrete indices \cite{lcrec, tiger}. However, these established tokenization strategies are fundamentally undifferentiated. They assign SIDs identically without accounting for inherent item popularity disparities, leading to unstructured and unpredictable branching points where head and tail item tokens compete.

\subsection{Popularity Bias and Fairness in Recommendation}
Traditional recommender systems frequently suffer from popularity bias, a phenomenon where a small fraction of highly interacted items disproportionately dominates algorithm exposure, leaving the long-tail of items largely ignored \cite{wsdm25bestpaper}. This algorithmic amplification, often referred to as the \textit{rich-get-richer} effect (a.k.a. the Matthew Effect), not only degrades user experience by failing to capture niche or diverse interests but also introduces critical fairness concerns across the platform, particularly for minority content creators who receive inequitable visibility. To mitigate these disparities, early literature primarily focused on heuristic post-processing and re-ranking techniques that explicitly boosted the exposure of long-tail items, albeit often at the expense of overall accuracy \cite{debias_survey}. More recently, the field has gravitated toward principled statistical and causal frameworks to achieve debiased learning. Schnabel et al. utilize inverse propensity weighting (IPW) \cite{ipw} to directly correct data collection biases by re-weighting user-item interactions during the model training phase. Furthermore, recent advances leveraging causal inference \cite{causal_inference_in_recsys} and adversarial learning have enabled systems to explicitly disentangle genuine user preferences from popularity-driven conformity, yielding robust representations that strive to optimize accuracy while maintaining multi-sided fairness for both users and providers \cite{causal_intervention}.

\subsection{Vector Quantization}
Vector Quantization (VQ) originated as a classical signal processing technique for data compression \cite{linde1980algorithm}, but it has recently become a cornerstone of deep representation learning. The seminal VQ-VAE \cite{vqvae} successfully integrated discrete latent spaces into neural architectures, effectively resolving the posterior collapse issue common in continuous models. Building upon this, frameworks like VQGAN \cite{vqgan} utilized adversarial objectives to dramatically enhance reconstruction fidelity. Crucially, by mapping high-dimensional data into discrete codebook indices, VQ allows complex signals to be treated as sequential tokens for autoregressive transformers. Beyond visual and audio synthesis, this discrete tokenization paradigm has recently advanced generative recommender systems. As we mentioned above, by employing VQ techniques, such as residual quantization \cite{rqkmeans,rqvae}, to discretize continuous item embeddings into semantic, categorical IDs, researchers have successfully reformulated recommendation as a sequence-to-sequence generation task \cite{tiger,lcrec,ed2}. This approach allows large language models to autoregressively predict next-item interactions using these discrete item tokens, seamlessly bridging traditional collaborative filtering with the powerful generative and reasoning capabilities of modern Transformer architectures.

\section{Experimental Detail}\label{app:detail}
\subsection{Dataset}\label{app:dataset}

\begin{table}[t]
  \caption{Statistics of the evaluated datasets. \textit{Avg.L} is the average length of the user interaction sequences.}
  \label{tab:dataset}
  \centering
  \resizebox{0.6\linewidth}{!}{
      \begin{tabular}{c|ccccc}
        \toprule
        \textbf{Dataset} & \textbf{\#User} & \textbf{\#Item} & \textbf{\#Interaction} & \textbf{Sparsity} & \textbf{Avg.L} \\
        \midrule
        Instruments & 24,772 & 9,922 & 206,153 & 99.92\% & 8.32 \\
        Games & 50,546 & 16,859 & 452,989 & 99.95\% & 8.96 \\
        Arts & 45,141 & 20,956 & 390,832 & 99.96\% & 8.66 \\
        \bottomrule
      \end{tabular}
  }
\end{table}

\begin{figure}[t]
    \centering
    \includegraphics[width=\textwidth]{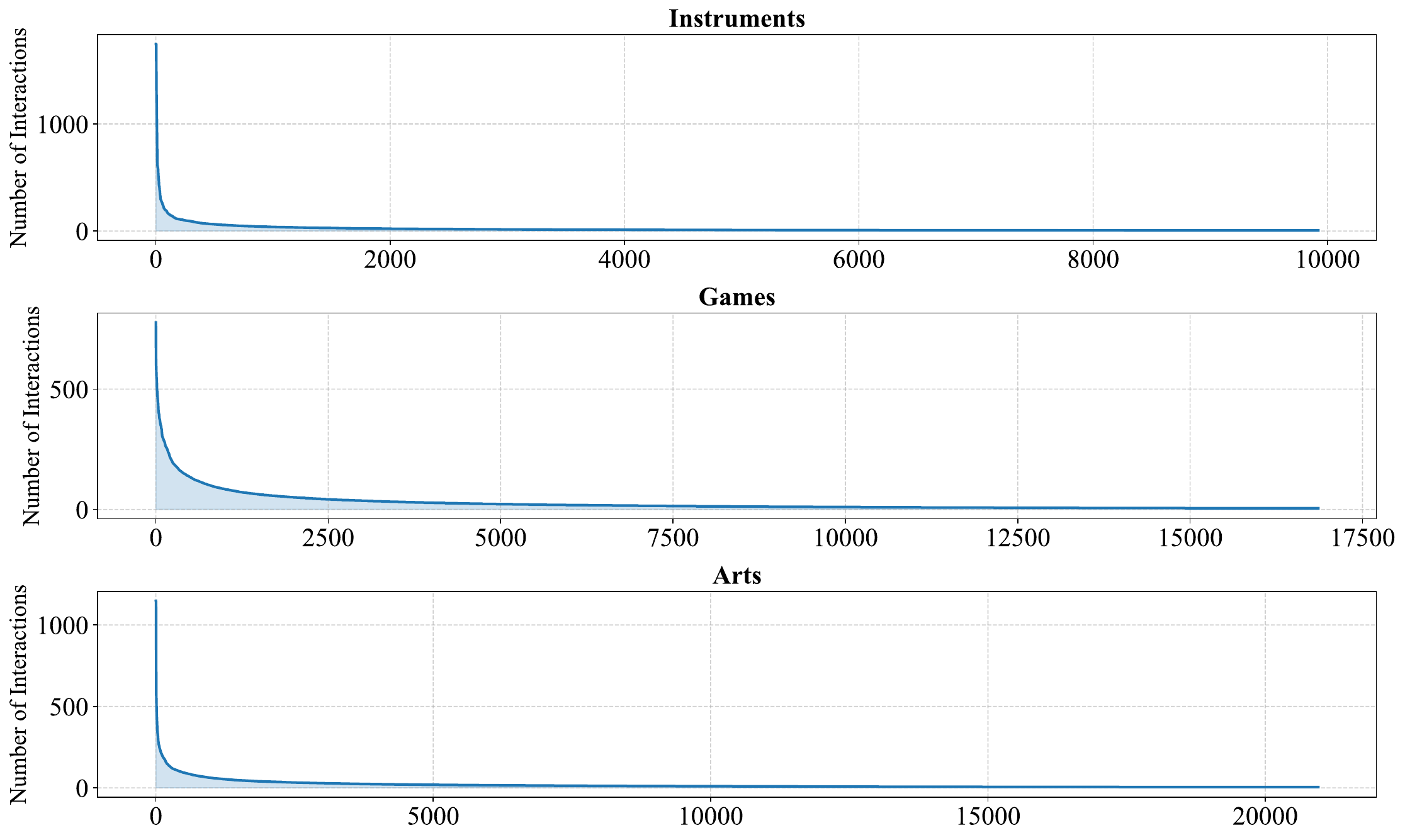}
    \caption{
    Long tail distribution of the item popularity.
    }
    \label{fig:long_tail}
    \vspace{-0.4cm}
\end{figure}

The dataset statistics are presented in Table \ref{tab:dataset}. The three sequential recommendation datasets originate from the Amazon Product Review dataset \cite{amazon_dataset}, which contains user review data from May 1996 to October 2018. Particularly, three categories for the sequential recommendation task, including \textit{"Musical Instruments"}, \textit{"Video Games"}, and \textit{"Arts, Crafts and Sewing"}, are extracted and organized into individual datasets \textit{Instruments}, \textit{Games}, and \textit{Arts}, respectively. Within the above datasets, each item is associated with a series of textual contents, including the item title, the detailed description, the item category, and so on. Similarly, the associated textual contents of the user entity include the user comment, the search query, and so on. Following standard procedure, inactive users/items with fewer than 5 interactions are filtered out, and the user interaction sequence is created in chronological order. In Figure \ref{fig:long_tail}, we present the distribution of item popularity for each of the three datasets, all of which are heavily long-talied.

\subsection{Baseline}\label{app:baseline}
$\bullet$ \textbf{LETTER} \cite{letter}, short for LEarnable Tokenizer for generaTivE Recommendation, is a learnable item tokenizer tailored for LLM-based generative recommendation. It addresses the limitations of existing ID, textual, and codebook-based identifiers by comprehensively integrating hierarchical semantics, collaborative signals, and code assignment diversity into item identifiers. Specifically, LETTER employs a RQ-VAE to encode item semantic information into hierarchical code sequences. To overcome the misalignment between semantics and collaborative signals, it introduces a contrastive alignment loss to align semantic quantized embeddings with collaborative filtering embeddings from well-trained models. Additionally, it applies a diversity loss based on constrained $K$-means clustering to regularize code embeddings, effectively mitigating code assignment and item generation biases. When instantiated on generative recommender models, LETTER further incorporates a ranking-guided generation loss to theoretically and empirically augment their top-$K$ ranking capabilities.

$\bullet$ \textbf{LC-Rec} \cite{lcrec} is a generative large language model based sequential recommendation approach designed to bridge the semantic gap between the language semantics of LLMs and the collaborative semantics of recommender systems. To represent items effectively without vocabulary explosion, LC-Rec employs a tree-structured vector quantization method based on item text embeddings to construct discrete item indices. It further utilizes a uniform semantic mapping technique \cite{sinkhorn} during index allocation to eliminate potential index conflicts among items. Instead of relying on predefined candidate sets, LC-Rec can autoregressively generate target items from the entire item set. To achieve deep integration of language and collaborative semantics, the LLM is fine-tuned on a series of specialized semantic alignment tasks, including sequential item prediction, explicit index-language alignment, and implicit recommendation-oriented alignment.

$\bullet$ \textbf{ED}$\mathbf{^2}$, i.e., the End-to-End Dual Dynamic recommender \cite{ed2}, is an LLM-based sequential recommender system that introduces a dual dynamic index mechanism. It addresses the limitations of existing LLM-based models that typically separate index generation from the sequential recommendation process and neglect user-related information. By utilizing a dual architecture with two homogeneous discrete index generators, ED$^2$ synchronously generates indices for both users and items, assembling index generation and sequential recommendation into a unified end-to-end LLM pipeline. To facilitate the LLM comprehension of the untrained dynamic index tokens, the model incorporates a multi-grained token regulator that establishes alignment supervision between dynamic index tokens and corresponding natural language tokens. Additionally, ED$^2$ leverages customized instruction tuning tasks and associated user collection data to exploit implicit high-order user-item interaction patterns based on historical behaviors.

$\bullet$ \textbf{\textit{Augmentation \& Substitution}} replace popular head items in the user interaction sequence with their most similar tail items according to a pre-defined replacement probability $p$. This straightforward principle aims to mitigate popularity bias by artificially increasing the representation of tail items during training. Specifically, for a typical long-tailed distribution where head items account for 80\% of total interactions and tail items account for 20\%, we set the target replacement probability $p=0.375$. This value is derived by solving the equilibrium equation $0.8-0.8p=0.2+0.8p$ to achieve a perfectly balanced ratio in the modified sequence. The target tail item for substitution is typically selected based on pre-calculated item similarity. Finally, the modified interaction sequence is integrated into the training process in two different ways. \textbf{\textit{(i)}} The \textit{Substitution} baseline directly overwrites the original sequence, achieving a strictly balanced head-to-tail ratio. \textbf{\textit{(ii)}} Alternatively, the \textit{Augmentation} baseline appends the modified sequence to the original training set. By combining the original data (0.8 head and 0.2 tail) with the modified data (0.5 head and 0.5 tail), \textit{Augmentation} yields a final head-to-tail ratio of 13:7, approximately 1.857, which significantly alleviates the initial 4:1 imbalance while strictly preserving the original interaction contexts.

$\bullet$ \textbf{IFairLRS} \cite{ifair} is an effective framework designed to enhance the item-side fairness of large language model-based recommendation systems (LRS). The framework addresses the item-side unfairness in LRS that primarily stems from two factors: the imbalanced distribution of historical user interactions and the inherent semantic biases present within LLMs. To mitigate these issues without sacrificing recommendation accuracy, IFairLRS calibrates recommendations by deploying two specifically adapted strategies across the main stages of building an LRS. During the instruction finetuning stage, it employs a reweighting strategy that adjusts the weights of training samples based on the bias observed between the distribution of target items and historical interactions. In the post learning stage (i.e., inference), it utilizes a re-ranking strategy that incorporates a punishment term based on group unfairness (GU) to adjust the final top-$K$ recommendations.

\subsection{Metric}
Following well-established benchmarks \cite{lcrec,handbook}, this study adopts Hit-Rate (HR) and normalized discounted cumulative gain (NDCG) to evaluate the recommendation performance. Formally, HR@$K$ is measured as follows,
\begin{equation}
    \text{HR}@K = \frac{1}{|U_{\text{test}}|} \sum_{u \in U_{\text{test}}} \mathbb{I}(\mathcal V_u \cap L_u^{(K)} \neq \emptyset),
\end{equation}
where $U_{\text{test}}$ denotes the user set for evaluation, $L_u^{(K)}$ denotes the top-$K$ recommendation list for user $u$, $\mathcal V_u$ represents the set of ground-truth interacted items for user $u$ in the test set, and $\mathbb{I}(\cdot)$ is an indicator function that returns $1$ if the condition is true and $0$ otherwise. 

Furthermore, NDCG@$K$ is defined below,
\begin{equation}
    \text{NDCG}@K = \frac{1}{|U_{\text{test}}|} \sum_{u \in U_{\text{test}}} \frac{\text{DCG}_u@K}{\text{IDCG}_u@K},
\end{equation}
\begin{equation}
    \text{DCG}_u@K=\sum_{i=1}^{K} \frac{r_{u,i}}{\log_2(i+1)}.
\end{equation}
Here, $r_{u,i} \in \{0, 1\}$ indicates whether the $i$-th recommended item in $L_u^{(K)}$ is relevant, i.e., exists in $\mathcal V_u$. $\text{IDCG}_u@K$ represents the ideal DCG score obtained by perfectly ranking all relevant items in $\mathcal V_u$ at the very top of the recommendation list.

This study adopts mean group unfairness (MGU) and average recommendation popularity (ARP) to quantify the fairness and the average popularity of the recommendation results \cite{2_8_split}. Particularly, MGU is measured as follows,
\begin{equation}
    \text{MGU} = (\text{GU}_{\text{head}}+\text{GU}_{\text{tail}})/2,
\end{equation}
where $\text{GU}_G$ stands for the unfairness of group $G$ and $G\in\{{\text{head}},{\text{tail}}\}$. 

$\text{GU}_G$ is defined as $\text{GU}_G=\text{GR}_G-\text{GH}_G$, where $\text{GR}_G$ and $\text{GH}_G$ represent the popularity of group $G$ in recommendation results and interaction history, respectively. Formally, let $\mathbf{H}$ denote the set of all user interaction sequences in the history, $\mathbf{L}$ denote the set of top-$K$ recommendations of all users at the inference phase, and $\mathcal{G}$ denote the set of item groups. Given an item group $G \in \mathcal{G}$, we can measure the recommendation proportion of group $G$ by
\begin{equation}
    \text{GR}_G = \frac{\sum_{L\in\mathbf{L}} \sum_{v\in L} \mathbb{I}(v \in G)}{\sum_{G'\in\mathcal{G}} \sum_{L\in\mathbf{L}} \sum_{v\in L} \mathbb{I}(v \in G')},
\end{equation}
where $\mathbb{I}(v \in G)$ is an identity function:
\begin{equation}
    \mathbb{I}(v \in G) = 
    \begin{cases} 
      1, & \text{item } v \text{ belongs to group } G \\
      0, & \text{otherwise} 
    \end{cases}.
\end{equation}
Intuitively, $\text{GR}_G$ calculates the recommendation proportion of group $G$ in the top-$K$ recommendations of all users. Accordingly, the interaction proportion of group $G$ in the historical interaction sequences $\mathbf{H}$ can be obtained by
\begin{equation}
    \text{GH}_G = \frac{\sum_{H \in \mathbf{H}} \sum_{v \in H} \mathbb{I}(v \in G)}{\sum_{G' \in \mathcal{G}} \sum_{H \in \mathbf{H}} \sum_{v \in H} \mathbb{I}(v \in G')}.
\end{equation}

Furthermore, ARP is defined below,
\begin{equation}
    \text{ARP}=\frac{1}{|U_{\text{test}}|}\sum_{u\in U_{\text{test}}}\frac{\sum_{i\in L_u}\varphi(i)}{|L_u|},
\end{equation}
where $U_{\text{test}}$ denotes the user set for evaluation, $L_u$ denotes the recommendation list for user $u$, and $\varphi(i)$ returns the popularity (i.e., the number of occurrences in the training set) of item $i$. 

\subsection{Implementation Detail}\label{app:imple_detail}

\begin{wraptable}{r}{0.3\textwidth}
  \caption{Hidden state dimensions of Qwen3 series.}
  \label{tab:qwen3_dim}
  \centering
  \resizebox{\linewidth}{!}{
      \begin{tabular}{c|c}
        \toprule
        \textbf{Scale} & \textbf{Hidden Dimension} \\
        \midrule
        0.6B & 1024 \\
        1.7B & 2048 \\
        4B & 2560 \\
        8B & 4096 \\
        \bottomrule
      \end{tabular}
  }
\end{wraptable}

This study employs the open source large language model \textit{Qwen} \cite{qwen2,qwen2.5,qwen3} developed by Alibaba. In particular, the experiments in Section \ref{sec:experiment}, including the main result, the ablation study, and the SID length analysis, adopt \href{https://modelscope.cn/models/Qwen/Qwen2.5-3B}{\textit{Qwen2.5-3B}} as the backbone model for all the evaluated GR models (i.e, Ghost, standard GRs, and GRs popularity debiasing methods). For the scaling pattern analysis, we adopt the fancy \href{https://modelscope.cn/collections/Qwen3-9743180bdc6b48}{\textit{Qwen3 series}}, and the backbone scale ranges from 0.6B to 8B. The hidden state dimension of Qwen2.5-3B is 2,048. The hidden state dimensions of the Qwen3 series are listed in Table \ref{tab:qwen3_dim}. The original size of all the Qwen vocabulary is 151,936.

\begin{wrapfigure}{r}{0.3\textwidth}
    \centering
    \includegraphics[width=0.95\linewidth]{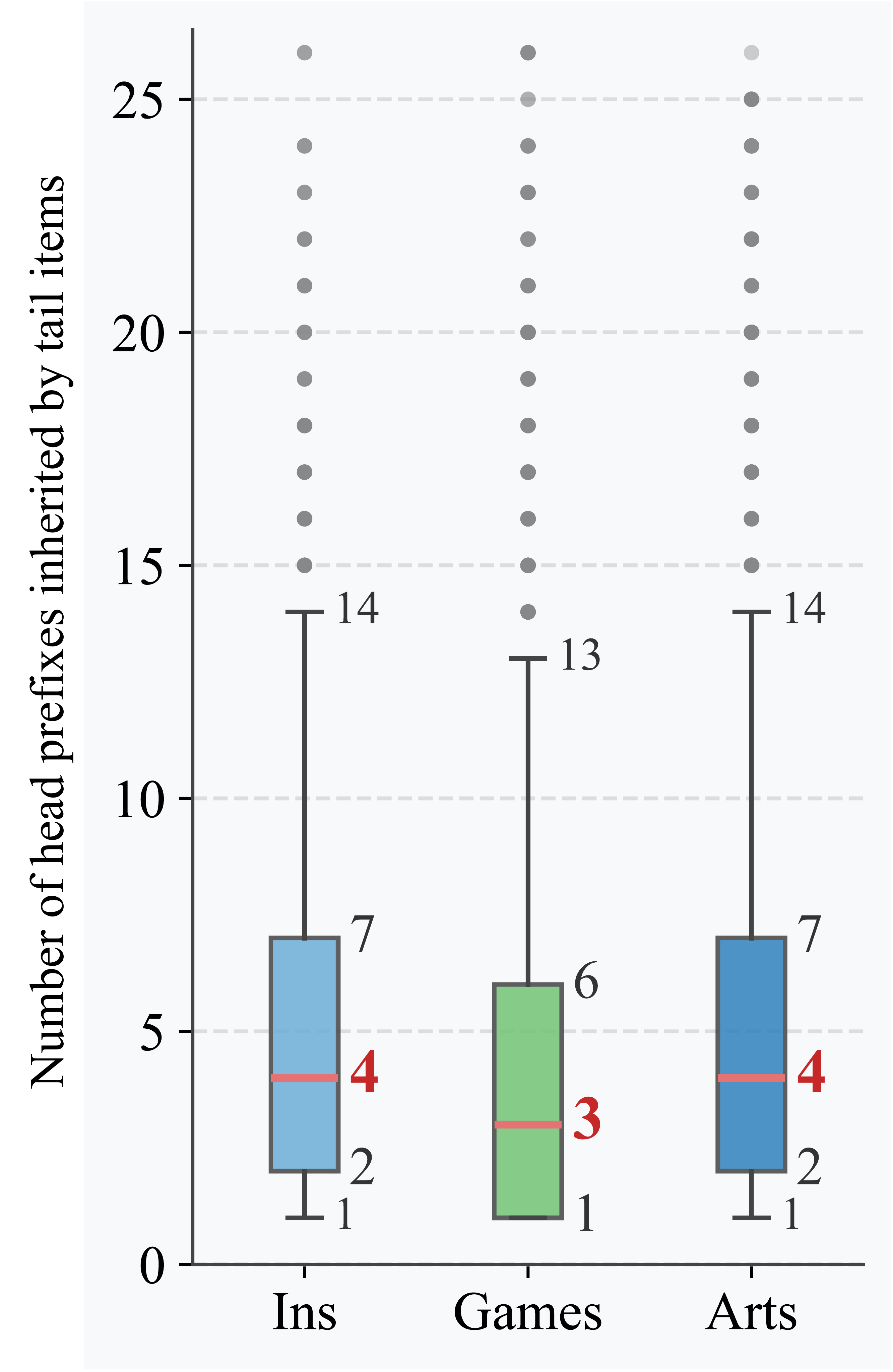}
    \caption{Number of tail items that inherit SID prefix from the same head items.}
    \label{fig:prefix_number}
\end{wrapfigure}

\textit{Within the skeleton-founded item tokenization (SKT)}, a series of linear layers [4096, 2048, 1024, 512, 256, 128, 64] is adopted to gradually reduce the Qwen representation into a 32-dimensional space for RQ-Kmeans. Within each RQ-Kmeans iteration, the number of clustering centers is 256 for both head items and tail items. For head items and tail items, the lengths of SIDs generated by SKT are distinct. The length of the head item SIDs exactly equals $L^h$, while that of the tail item SIDs is $L^h+L^t$. Hence, the SIDs generated by SKT are essentially indices with variable lengths. Figure \ref{fig:prefix_number} presents a box plot illustrating the distribution of the number of head prefixes inherited by tail items across three datasets. The visualization highlights the central tendency, variance, and outliers within each domain. The data indicate that the Ins and Arts datasets share identical descriptive statistics, both demonstrating a median value of 4 and an interquartile range (IQR) spanning from 2 to 7. However, the Games dataset exhibits slightly lower overall values, featuring a median of 3 and an IQR ranging from 1 to 6.
\textit{For the asymmetric unlikelihood optimization (AUO)}, the default values of $K_a,K_b$ controlling the candidate scale of the undesired item selection are set to 200 and 5. Therefore, for each tail item $v'$, the cardinality of the undesired collection $\mathbf{\mathbf{\Omega}}_{v'}$ is 5. Regarding the \model training phase, we adopt the AdamW optimizer \cite{adamw} with a learning rate of $3\times10^{-5}$. In Section \ref{app:hyper}, we conduct an analysis of the undesired collection size and the learning rate. All the experiments are completed on a machine with 8 \textit{NVIDIA A100 Tensor Core 80GB} GPUs.

\section{Equal-sized Grouping based on Item Popularity}\label{app:equal_size}

\begin{figure}[t]
    \centering
    \includegraphics[width=\textwidth]{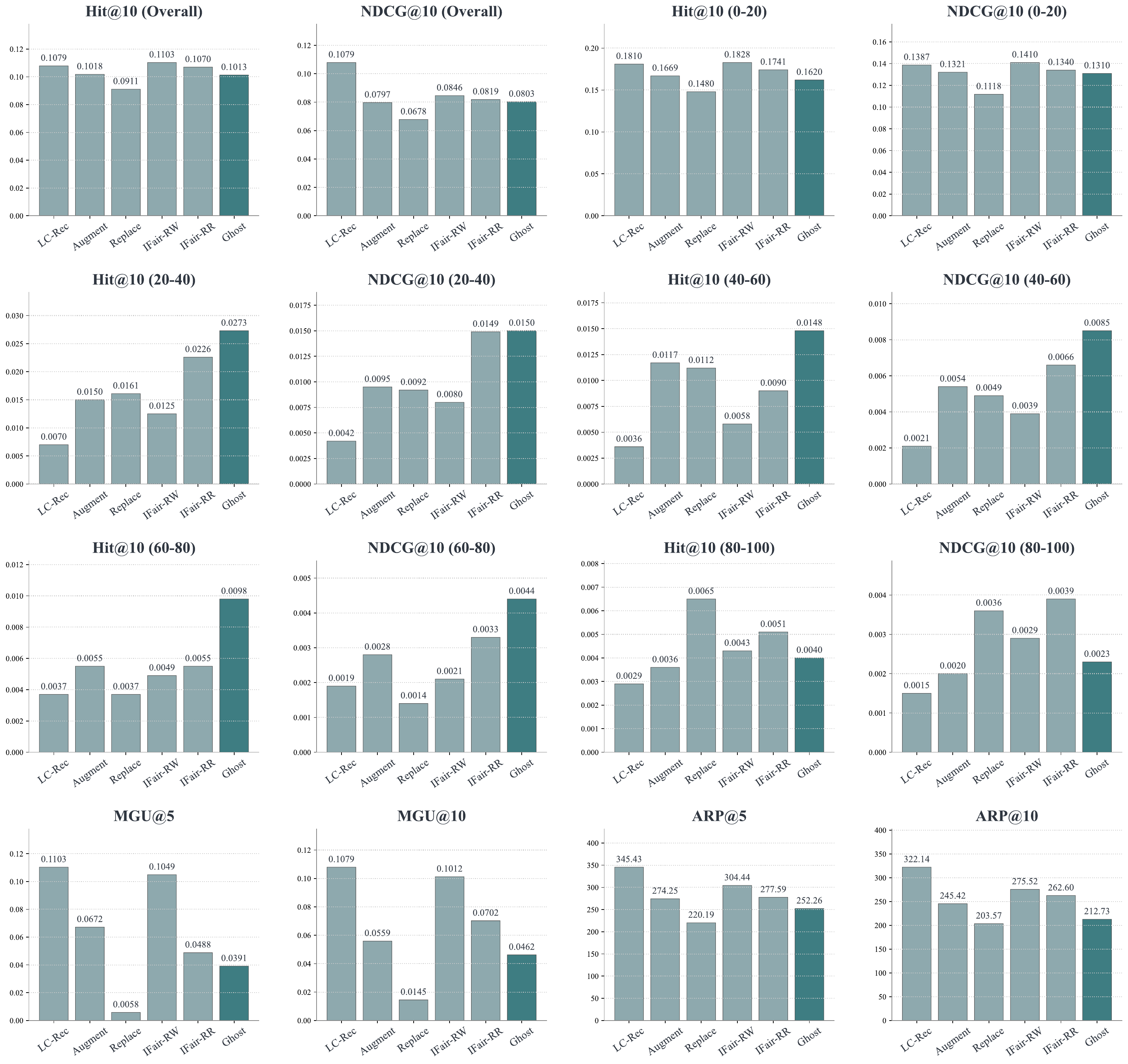}
    \caption{
    Performance comparison of each equal-sized grouping on Ins dataset.
    }
    \label{fig:equal_group}
    \vspace{-0.2cm}
\end{figure}

To provide a more fine-grained understanding of where the performance improvements originate, we analyze the recommendation results across different popularity segments. Figure \ref{fig:equal_group} details the performance comparison on the Ins dataset, breaking down the item space into five equal-sized groups sorted by popularity (from the most popular 0-20 segment to the least popular 80-100 segment), alongside overall accuracy and debiasing metrics. In general, the fine-grained results confirm that Ghost successfully shifts exposure from over-recommended head items to underexposed mid-tail and tail items without sacrificing general utility. Specifically, we can draw the following three insights.

\textbf{\textit{First}}, Ghost demonstrates an exceptional capability to excavate and accurately recommend items across the broad long-tail distribution. For the intermediate and long-tail segments (specifically the 20-40, 40-60, and 60-80 groups), Ghost consistently achieves the highest Hit@10 and NDCG@10 scores among all evaluated models. For example, in the 20-40 group, Ghost reaches a Hit@10 of 0.0273, substantially outperforming the standard LC-Rec baseline (0.0070) and the strongest fairness-aware baseline, IFair-RR (0.0226). This proves the model's targeted effectiveness in surfacing relevant, less-mainstream content. \textbf{\textit{Second}}, Ghost effectively suppresses the over-exposure of head items, leading to significantly enhanced recommendation fairness. This is clearly evidenced by the steep reductions in Average Recommendation Popularity (ARP) and Monopoly Gini of Users (MGU) metrics. Ghost lowers the ARP@10 to 212.73 and MGU@10 to 0.0462, representing a massive drop compared to LC-Rec's 322.14 and 0.1079, respectively. While the heuristic "Replace" baseline achieves slightly lower ARP and MGU scores, it does so by indiscriminately swapping items, which severely damages recommendation quality.

\section{Supplement Hyper-parameter Analysis}

\begin{figure}[t]
    \centering
    \includegraphics[width=\textwidth]{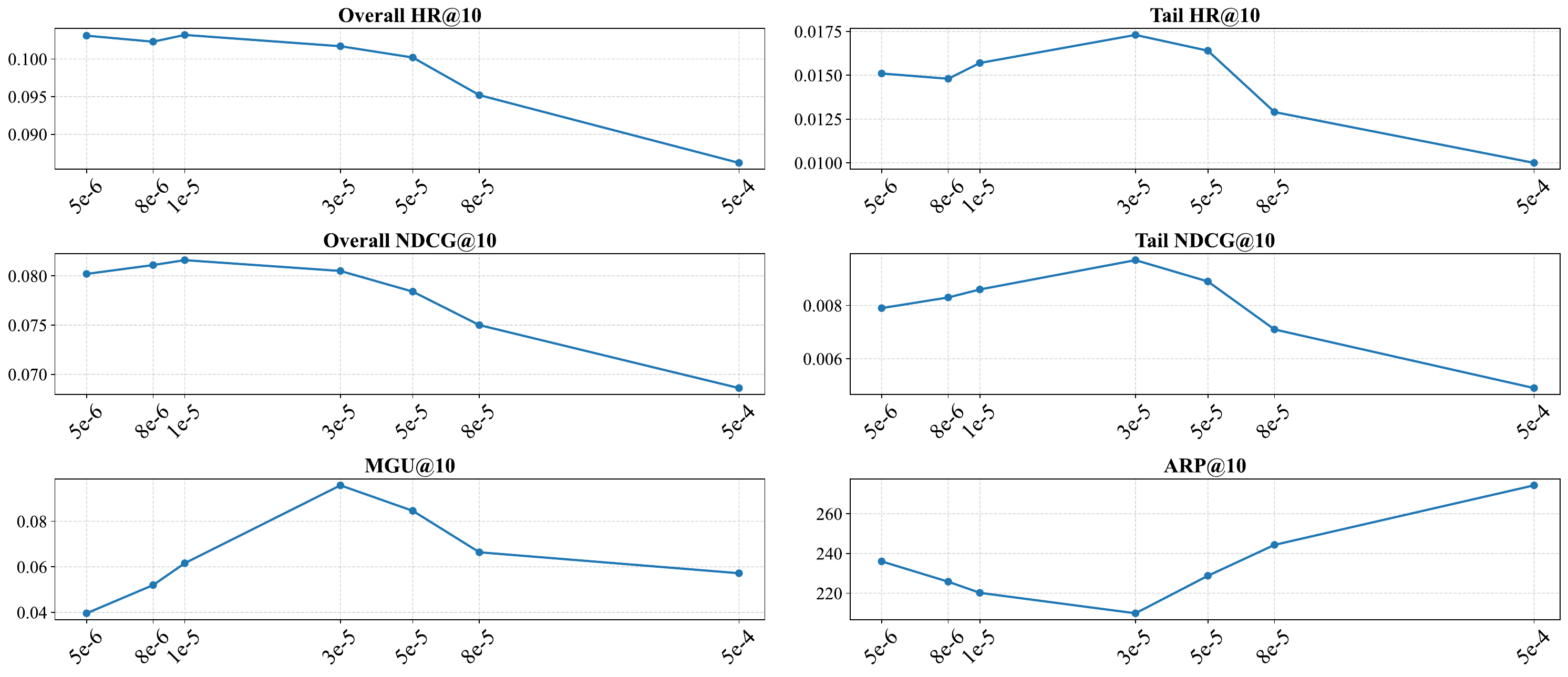}
    \caption{
    Tendency of \model performance on Ins dataset, under different learning rates. The $x$-axis denotes learning rate, and the $y$-axis is the metric values.
    }
    \label{fig:lr}
    \vspace{-0.2cm}
\end{figure}

\subsection{Learning rate}
Here, we investigate the impact of the learning rate during \model optimization. Figure \ref{fig:lr} illustrates the sensitivity of the Ghost model to varying learning rates on the Ins dataset, revealing its crucial role in balancing overall recommendation accuracy with popularity bias mitigation. The empirical trends identify $3\times 10^{-5}$ as the optimal learning rate configuration. At this specific point, the model attains peak effectiveness in long-tail item excavation, evidenced by the highest Tail HR@10 and Tail NDCG@10, while simultaneously maximizing user coverage (MGU@10). Crucially, the ARP@10 reaches its global minimum at this setting, indicating a strong suppression of popularity bias without severely compromising the highly competitive overall accuracy metrics (Overall HR@10 and NDCG@10, which peak slightly earlier around $1\times 10^{-5}$). Conversely, employing an excessively large learning rate (e.g., $5\times 10^{-4}$) yields detrimental effects across all dimensions: overall and tail-specific accuracies sharply degrade, user coverage diminishes, and ARP@10 spikes significantly. This demonstrates that an overly aggressive learning rate destabilizes the debiasing mechanism, causing the model to collapse back into disproportionately favoring mainstream items.

\begin{figure}[t]
    \centering
    \includegraphics[width=\textwidth]{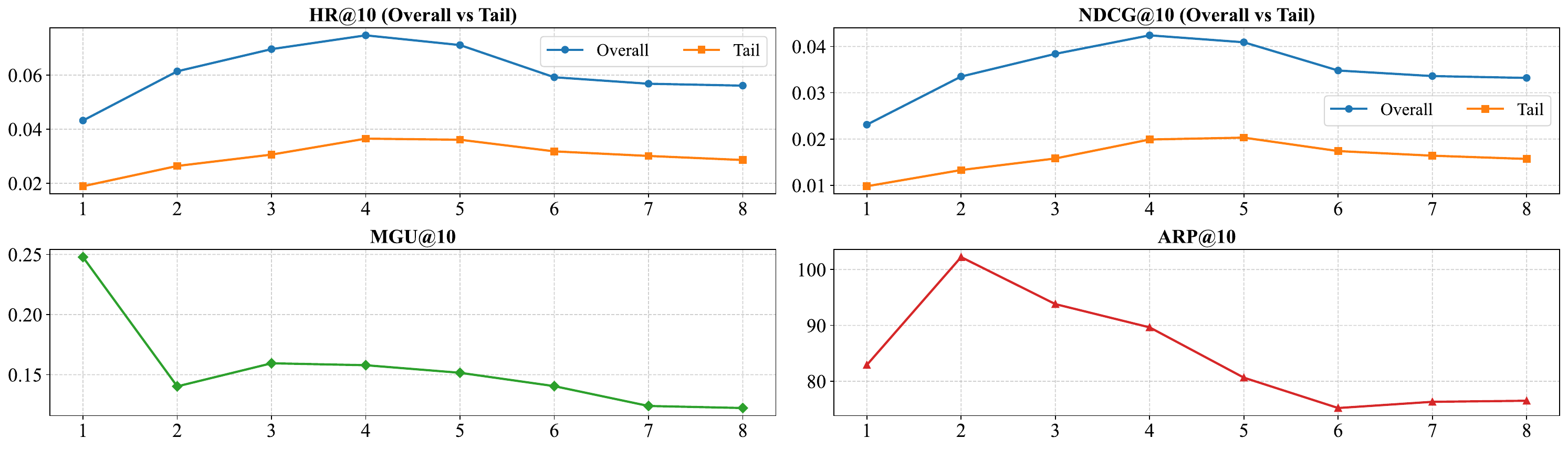}
    \caption{
    Tendency of \model performance on Games dataset, under different optimization epochs. The $x$-axis denotes the epoch values, and the $y$-axis is the metric values.
    }
    \label{fig:epoch}
    \vspace{-0.4cm}
\end{figure}

\subsection{Epoch}
Here, we investigate the impact of the optimization epoch. Figure \ref{fig:epoch} visualizes the impact of the number of optimization epochs on the Ghost model performance using the Games dataset. The empirical trajectories indicate that determining an appropriate stopping point is crucial for balancing recommendation accuracy with popularity bias mitigation. Specifically, epoch 4 emerges as the optimal training duration, where both overall and tail-specific accuracy metrics (HR@10 and NDCG@10) simultaneously achieve their peak performance. During the initial training phase (epochs 1 and 2), the model exhibits suboptimal accuracy and a sharp spike in Average Recommendation Popularity (ARP@10), despite an initially high user coverage (MGU@10) that quickly drops. As optimization progresses towards epoch 4, the model effectively converges, significantly enhancing long-tail item retrieval while steadily driving down ARP@10. Conversely, prolonged training (epochs 6 through 8) evidently leads to overfitting; this over-optimization results in a consistent degradation across both overall and tail accuracies, a continued decline in MGU@10, and yields no further meaningful benefits for popularity debiasing. Therefore, maintaining a moderate number of optimization epochs is essential to maximize the framework's capacity to deliver accurate, balanced, and diverse recommendations.

\section{Theoretical Derivations and Justifications}\label{app:theory}

This section provides the rigorous theoretical foundations and detailed mathematical proofs for the lemmas and corollaries presented in the main text regarding the Ghost model. Specifically, it elucidates the gradient starvation issue inherent in Maximum Likelihood Estimation (MLE), the localized bias amplification effect induced by undifferentiated tokenization, and the mathematical mechanisms through which Skeleton-Founded Tokenization (SKT) and Asymmetric Unlikelihood Optimization (AUO) structurally mitigate popularity bias.

\subsection{Prerequisites and Theoretical Foundations}

To establish a rigorous theoretical framework for diagnosing popularity bias in Generative Recommenders (GRs), we first formalize three fundamental prerequisites regarding the data distribution and the architecture of SID-based GRs.

\begin{itemize}
[leftmargin=*]
\item \textbf{Prerequisite 1 (Long-tail Distribution)}. In the sequential recommendation training dataset $\mathcal{D}$, the frequency of user-item interactions exhibits a heavily skewed, long-tailed distribution (e.g., a power-law distribution). Consequently, the occurrence probability of items in the head set $\mathcal{V}_{\text{head}}$ (the top 20\% most popular items) is vastly greater than that of items in the tail set $\mathcal{V}_{\text{tail}}$ (the remaining 80\%), i.e., $\mathbb{P}_{\mathcal{D}}(v \in \mathcal{V}_{\text{head}}) \gg \mathbb{P}_{\mathcal{D}}(v \in \mathcal{V}_{\text{\text{tail}}})$.
    
\item \textbf{Prerequisite 2 (Inner Product Mapping)}. Current SOTA GR architectures formulate the next-token prediction task by computing generative logits via the inner product of the user historical behavior representation $X_{h_u}$ and the candidate token embedding $e_c$. The generation probability is formalized via a Softmax operation across the token vocabulary,
\begin{equation}
    \mathcal{P}_\theta(c|h_u) = \frac{\exp(\langle e_c, X_{h_u} \rangle)}{\sum_{c'} \exp(\langle e_{c'}, X_{h_u} \rangle)} \label{eq:softmax_assump}
\end{equation}
    
\item \textbf{Assumption 1 (Identical Token Space under Undifferentiated Tokenization)}. Current vector quantization techniques (e.g., RQ-VAE, RQ-KMeans) process head and tail items indiscriminately, meaning they share and compete within the identical representation space. Let $c_{\text{tail}}$ denote a token structurally exclusive to tail items. Due to the heavy-tailed interaction distribution established in Prerequisite 1, the mathematical expectation of $c_{\text{tail}}$ acting as a ground-truth target token within the training distribution $\mathcal{D}$ asymptotically approaches zero.
\end{itemize}

\subsection{Detailed Derivations of LEMMA 1 and COROLLARY 1}

\textbf{Proof of LEMMA 1 (Gradient Starvation in MLE)}.

\textit{Basically}, the model is optimized using standard Maximum Likelihood Estimation, specifically minimizing the Negative Log-Likelihood loss over the sequential tokens of the target item SID.

For an arbitrary generation step $i$ and target item $v$, the instantaneous loss is $$\mathcal{L}_{\text{NLL}}^{(i)} = -\log \mathcal{P}_\theta(c_v^{(i)}|h_u, c_v^{<i}).$$ According to the standard derivative properties of the Softmax function in Eq.\eqref{eq:softmax_assump}, the partial derivative of the prediction probability with respect to an arbitrary candidate token embedding $e_c$ yields a Jacobian formulation that separates into target and non-target cases. Consequently, the gradient of the NLL loss with respect to $e_c$ is,
\begin{equation}
    \frac{\partial \mathcal{L}_{\text{NLL}}^{(i)}}{\partial e_c} = \left( \mathcal{P}_\theta(c|h_u, c_v^{<i}) - \mathbb{I}\{c = c_v^{(i)}\} \right) X_{h_u} \label{eq:nll_grad}
\end{equation}

In gradient descent, the parameter update direction opposes the computed gradient, denoted as $\Delta e_c \propto -\frac{\partial \mathcal{L}}{\partial e_c}$. Taking the mathematical expectation over the entire training distribution $\mathcal{D}$, we obtain,
\begin{equation}
    \mathbb{E}_{\mathcal{D}}[\Delta e_c] \propto \mathbb{E}_{\mathcal{D}}\left[ \sum_i \left( \mathbb{I}\{c = c_v^{(i)}\} - \mathcal{P}_\theta(c|h_u, c_v^{<i}) \right) X_{h_u} \right] \label{eq:expected_update}
\end{equation}

Now, consider a tail-specific token $c_{\text{tail}}$. Based on Prerequisite 1 and Assumptions 1, its probability of being sampled as the ground-truth target token within any batch drawn from $\mathcal{D}$ is negligible, $\mathbb{P}_{\mathcal{D}}(c_{\text{tail}} = c_v^{(i)}) \approx 0$. Consequently, the indicator function term $\mathbb{I}\{c_{\text{tail}} = c_v^{(i)}\}$ reliably vanishes. By projecting the expected gradient update onto the user preference vector $X_{h_u}$, we isolate the token alignment trajectory, formulated as,
\begin{equation}
    \mathbb{E}_{\mathcal{D}}[\langle \Delta e_{c_{\text{tail}}}, X_{h_u} \rangle] \approx -\mathbb{E}_{\mathcal{D}}\left[ \sum_i \mathcal{P}_\theta(c_{\text{tail}}|h_u, c_v^{<i}) \cdot \|X_{h_u}\|_2^2 \right] \le 0 \label{eq:starvation}
\end{equation}

\textit{\textbf{Conclusion}}. Eq.\eqref{eq:starvation} rigorously proves that tail tokens consistently act as trivial negative samples within the Softmax denominator during MLE optimization. Lacking positive reinforcement from the indicator function, their embeddings $e_{c_{\text{tail}}}$ are pathologically pushed away from the user intent space, trapping them in a state of Gradient Starvation \cite{gradient_starvation}.

\vspace{1em}
\textbf{Proof of COROLLARY 1 (Head Token Dominance at Branching Point)}.

Building upon LEMMA 1, the parameter space is assumed to have converged to a skewed state where head tokens have received massive positive updates due to high interaction frequencies, i.e., $\mathbb{I}\{c_{\text{head}} = c_v^{(i)}\} = 1$ frequently, while tail tokens are starved.

This optimization skew manifests in the representation space as an extreme disparity in inner products, $\langle e_{c_{\text{head}}}, X_{h_u} \rangle \gg \langle e_{c_{\text{tail}}}, X_{h_u} \rangle$. Let $i$ denote an arbitrary generative branching point where head and tail candidate tokens structurally compete. The predicted probability ratio for these tokens is given by,
\begin{equation}
    \frac{\mathcal{P}_\theta(c_{\text{head}}^{(i)}|h_u, c^{<i})}{\mathcal{P}_\theta(c_{\text{tail}}^{(i)}|h_u, c^{<i})} = \frac{\exp(\langle e_{c_{\text{head}}^{(i)}}, X_{h_u} \rangle)}{\exp(\langle e_{c_{\text{tail}}^{(i)}}, X_{h_u} \rangle)} = \exp\left( \langle e_{c_{\text{head}}^{(i)}} - e_{c_{\text{tail}}^{(i)}}, X_{h_u} \rangle \right) \label{eq:ratio}
\end{equation}

Because the gradient updates heavily favor the head token, the exponential term in Eq.\eqref{eq:ratio} is pathologically amplified. This causes the predictive distribution $\mathcal{P}_\theta$ to diverge significantly from the true underlying data distribution $\mathcal{P}_d$. We formally define this local divergence as the amplification factor $\gamma_i$ as follows,
\begin{equation}
    \gamma_i = \frac{\mathcal{P}_\theta(c_{\text{head}}^{(i)}|h_u, c^{<i}) / \mathcal{P}_\theta(c_{\text{tail}}^{(i)}|h_u, c^{<i})}{\mathcal{P}_d(c_{\text{head}}^{(i)}|h_u, c^{<i}) / \mathcal{P}_d(c_{\text{tail}}^{(i)}|h_u, c^{<i})} > 1 \label{eq:gamma}
\end{equation}
Eq.\eqref{eq:gamma} demonstrates the inevitable probability dominance of head tokens, establishing that the generative process becomes overconfident in predicting head tokens, regardless of context.

\subsection{Detailed Derivations of LEMMA 2 and LEMMA 3}

\textbf{Proof of LEMMA 2 (Bias Amplification via Undifferentiated Tokenization)}.

\textit{Basically}, the current item tokenization method assigns Semantic Indices (SIDs) of length $L$ indiscriminately, treating all items equally without accounting for popularity.

Under undifferentiated tokenization, a long-tail item $v_{\text{tail}}$ shares prefixes of varying lengths with numerous head items. Consequently, generating $v_{\text{tail}}$ requires decoding a sequence of tokens where it must repeatedly survive competition against popular items. 

Suppose there exists a set $\mathcal Z$ containing $z$ unpredictable branching points ($|\mathcal Z| = z \le L$) where tail tokens structurally compete against head tokens. Based on the chain rule of autoregressive sequence generation and the dominance established in Corollary 1, each time the generation navigates a competitive branching point $j \in Z$, the relative generation probability of the tail token is suppressed by at least a factor of $\gamma_j$,
\begin{equation}
    \mathcal{P}_\theta(c_{\text{tail}}^{(j)}|h_u, c^{<j}) \le \gamma_j^{-1} \cdot \mathcal{P}_d(c_{\text{tail}}^{(j)}|h_u, c^{<j}) \label{eq:step_suppression}
\end{equation}

Over the entire generation sequence of $v_{\text{tail}}$, we multiply the conditional probabilities. The suppressions at branching points accumulate geometrically,
\begin{align}
    \mathcal{P}_\theta(v_{\text{tail}}|h_u) &= \prod_{j=1}^L \mathcal{P}_\theta(c_{\text{tail}}^{(j)}|h_u, c^{<j}) \nonumber \\
    &\le \left( \prod_{j \in Z} \gamma_j^{-1} \right) \prod_{j=1}^L \mathcal{P}_d(c_{\text{tail}}^{(j)}|h_u, c^{<j}) \nonumber \\
    &\le (\gamma_{\text{min}})^{-z} \prod_{j=1}^L \mathcal{P}_d(c_{\text{tail}}^{(j)}|h_u, c^{<j}) \label{eq:geometric_suppression}
\end{align}
where $\gamma_{\text{min}} = \min_{j \in \mathcal Z}(\gamma_j) > 1$.

\textit{\textbf{Conclusion}}. This derivation establishes that undifferentiated tokenization cascades localized, token-level gradient bias into a macroscopic, geometric $\mathcal{O}(\gamma_{\text{min}}^z)$ probability suppression for tail items, fully explaining their severe marginalization in recommendation lists.

\vspace{1em}
\textbf{Proof of LEMMA 3 (Mitigation of Bias Amplification via SKT)}.

The Skeleton-Founded Tokenization (SKT) mechanism asynchronously defines SIDs. The head items dictate the $L^h$-length skeleton of the SID space. A tail item $v'$ is forced to explicitly inherit the first $L^h$ tokens from its semantically closest head item $v^*$, and subsequently generates $L^t$ specific tokens to characterize its distinctiveness.

During the initial generation steps $j \in [1, L^h]$, the tail item $v'$ and the head item $v^*$ share an identical prefix trajectory ($c_{v'}^{(j)} = c_{v^*}^{(j)}$). Consequently, no probability divergence or chaotic token competition occurs between them. The unstructured branching points that comprised set $\mathcal Z$ in LEMMA 2 are entirely eliminated. The genuine structural divergence is uniformly deferred to the single, predictable step $(L^h + 1)$. At this exact locus, the head item outputs an End-Of-Sequence (EOS) token, whereas the tail item generates the first token of its distinct semantic prefix. Because the competition is now strictly restricted to this single step, the cardinality of the branching set $z$ is explicitly limited to 1. 

The geometric suppression series detailed in Eq.\eqref{eq:geometric_suppression} therefore collapses into a singular, localized deviation bounded by the head-dominance factor at the EOS step ($\gamma_{EOS}$):
\begin{equation}
    \mathcal{P}_\theta(v'|h_u) = \prod_{j=1}^{L^h+L^t} \mathcal{P}_\theta(c_{v'}^{(j)}|h_u, c^{<j}) \approx (\gamma_{EOS})^{-1} \prod_{j=1}^{L^h+L^t} \mathcal{P}_d(c_{v'}^{(j)}|h_u, c^{<j}) \label{eq:collapsed_bias}
\end{equation}

\textit{\textbf{Conclusion}}. This rigorously proves that SKT effectively halts the Markovian amplification chain of popularity bias. By establishing a unified structural branching point, it transforms a multi-step geometric suppression $\mathcal{O}(\gamma_{\text{min}}^z)$ into an insulated, single-step discrepancy $\mathcal{O}(\gamma_{EOS})$.

\subsection{Detailed Derivations of LEMMA 4}

\textbf{Proof of LEMMA 4 (Gradient Rescue based on AUO)}.

The \model model incorporates Asymmetric Unlikelihood Optimization (AUO) to actively penalize a dynamically generated set of undesired tokens $\bar{\mathbf\Omega}$. For a target tail item $v'$, $\bar{\mathbf\Omega}$ consists of SIDs from head items that share high textual similarity with $v'$ but possess divergent SID structures, acting as deceptive popular distractions.

The AUO loss function introduces an explicit penalty for generating tokens in $\bar{\mathbf\Omega}$,
\begin{equation}
    \mathcal{L}_{\text{AUO}} = -\sum_{c \in \bar{\mathbf\Omega}} \log(1 - \mathcal{P}_\theta(c))
\end{equation}

To derive the partial derivative with respect to an arbitrary candidate embedding $e_k$, let the logit be defined as $z_k = \langle e_k, X_{h_u} \rangle$. We first utilize the Jacobian matrix of the Softmax function for individual output probabilities,
\begin{itemize}
[leftmargin=*]
    \item For the diagonal term ($i = k$), $$\frac{\partial \mathcal{P}_\theta(c_i)}{\partial z_k} = \mathcal{P}_\theta(c_i)(1 - \mathcal{P}_\theta(c_i))$$
    \item For the off-diagonal terms ($i \neq k$), $$\frac{\partial \mathcal{P}_\theta(c_i)}{\partial z_k} = -\mathcal{P}_\theta(c_i)\mathcal{P}_\theta(c_k)$$
\end{itemize}

Applying the chain rule to the AUO objective yields:
\begin{equation}
    \frac{\partial \mathcal{L}_{\text{AUO}}}{\partial z_k} = \sum_{c_i \in \bar{\mathbf\Omega}} \frac{1}{1 - \mathcal{P}_\theta(c_i)} \frac{\partial \mathcal{P}_\theta(c_i)}{\partial z_k} \label{eq:auo_chain}
\end{equation}

\textbf{Scenario 1: For a false positive head token $c_{\text{head}}^- \in \bar{\mathbf\Omega}$ (Active Suppression)}.

Here, the embedding of interest $k = c_{\text{head}}^-$ belongs to the actively penalized set. We separate the summation in Eq.\eqref{eq:auo_chain} into terms where $i=k$ and $i \neq k$,
\begin{equation}
    \frac{\partial \mathcal{L}_{\text{AUO}}}{\partial z_{\text{head}}^-} = \frac{\mathcal{P}_\theta(c_{\text{head}}^-)(1 - \mathcal{P}_\theta(c_{\text{head}}^-))}{1 - \mathcal{P}_\theta(c_{\text{head}}^-)} + \sum_{c_i \in \bar{\mathbf\Omega} \setminus \{c_{\text{head}}^-\}} \frac{-\mathcal{P}_\theta(c_i)\mathcal{P}_\theta(c_{\text{head}}^-)}{1 - \mathcal{P}_\theta(c_i)} \label{eq:scenario1_split}
\end{equation}
Simplifying the first term, we get,
\begin{equation}
    \frac{\partial \mathcal{L}_{\text{AUO}}}{\partial z_{\text{head}}^-} = \mathcal{P}_\theta(c_{\text{head}}^-) - \sum_{c_i \neq c_{\text{head}}^-} \frac{\mathcal{P}_\theta(c_i)\mathcal{P}_\theta(c_{\text{head}}^-)}{1 - \mathcal{P}_\theta(c_i)} \label{eq:scenario1_simp}
\end{equation}
The overall objective is a weighted combination of MLE and AUO ($\mathcal{L} = \mathcal{L}_{NLL} + \alpha \mathcal{L}_{\text{AUO}}$). Incorporating the standard MLE base gradient (which negatively updates $c_{\text{head}}^-$ as it is a false negative, not the target), the overall parameter update direction becomes,
\begin{equation}
    \Delta e_{c_{\text{head}}^-} \propto -(1+\alpha)\mathcal{P}_\theta(c_{\text{head}}^-)X_{h_u} + \alpha \sum_{c_j \in \bar{\mathbf\Omega} \setminus \{c_{\text{head}}^-\}} \frac{\mathcal{P}_\theta(c_j)\mathcal{P}_\theta(c_{\text{head}}^-)}{1-\mathcal{P}_\theta(c_j)} X_{h_u} \label{eq:cross_penalization}
\end{equation}
This explicitly shows the direct penalty via the softmax derivative.

\textbf{Scenario 2: For a target tail token $c_{\text{tail}} \notin \bar{\mathbf\Omega}$ (Cross-Penalization Rescue)}.

Here, the embedding of interest $k = c_{\text{tail}}$ corresponds to our ground-truth tail item, which is specifically excluded from the explicitly penalized head set $\bar{\mathbf\Omega}$. Therefore, the condition $i \neq k$ strictly holds for \textit{all} terms in the summation of Eq.\eqref{eq:auo_chain},
\begin{equation}
    \frac{\partial \mathcal{L}_{\text{AUO}}}{\partial z_{\text{tail}}} = \sum_{c_j \in \bar{\mathbf\Omega}} \frac{-\mathcal{P}_\theta(c_j)\mathcal{P}_\theta(c_{\text{tail}})}{1 - \mathcal{P}_\theta(c_j)} \label{eq:scenario2_grad}
\end{equation}
In gradient descent, the specific parameter update contribution derived from AUO towards this tail token acts in the opposing direction ($-\alpha\cdot\partial \mathcal{L}_{\text{AUO}}/{\partial z_{\text{tail}}}$),
\begin{equation}
    \Delta e_{c_{\text{tail}}}^{\text{AUO}} \propto +\alpha \sum_{c_j \in \bar{\mathbf\Omega}} \frac{\mathcal{P}_\theta(c_j)\mathcal{P}_\theta(c_{\text{tail}})}{1 - \mathcal{P}_\theta(c_j)} X_{h_u} \label{eq:auo_positive_force}
\end{equation}
Merging this positive AUO rescue term with the suppressive MLE base term established in LEMMA 1 rigorously completes the derivation:
\begin{equation}
    \Delta e_{c_{\text{tail}}} \propto -\mathcal{P}_\theta(c_{\text{tail}}) X_{h_u} + \alpha \sum_{c_j \in \bar{\mathbf\Omega}} \frac{\mathcal{P}_\theta(c_j)\mathcal{P}_\theta(c_{\text{tail}})}{1 - \mathcal{P}_\theta(c_j)} X_{h_u} \label{eq:rescue_complete}
\end{equation}

\textit{\textbf{Conclusion}}. This Jacobian analysis illuminates how the AUO loss function systematically transfigures the penalization of notorious head tokens into a positive, structural rescue force for long-tail tokens. By redistributing probability mass, this mechanism actively counteracts and dismantles the gradient starvation trap established in LEMMA 1, ensuring tail tokens receive rational parameter updates.

\section{Limitations and Future Work}\label{app:limitation}

$\bullet$ \textbf{Supervised Finetuning}. While this study provides a comprehensive diagnosis and mitigation strategy for popularity bias in GRs, the scope of our investigation is currently constrained to the supervised fine-tuning (SFT) paradigm. Specifically, our theoretical analysis of the gradient starvation issue and the subsequent design of the AUO are fundamentally grounded in the MLE framework. Recent advancements in generative modeling have increasingly adopted reinforcement learning (RL) techniques to align models with complex objectives \cite{openonerec,sprec}. The specific mechanisms by which popularity bias manifests in RL-based GRs, and whether our proposed skeleton-founded tokenization and unlikelihood penalties remain effective under reward-driven optimization, have not yet been analyzed and represent a critical direction for future research.

$\bullet$ \textbf{Pre-computed Undesired Collection}. The undesired collection utilized in the AUO module relies on a static, pre-computed approach. As detailed in our methodology, the undesired items are identified by retrieving head items with high initial semantic similarity but divergent SIDs relative to the target tail item. Currently, this collection is established prior to training and is not dynamically adjusted as the model's parameters and internal representations evolve during the optimization process. While dynamically updating the undesired collection at each epoch could theoretically provide more precise and adaptive negative supervision, the static approach was deliberately adopted as a pragmatic trade-off to preserve computational efficiency and prevent prohibitive training overhead.

$\bullet$ \textbf{Online A/B Test}. At present, deploying and comprehensively evaluating the proposed \model model within live, large-scale industrial systems presents objective difficulties for us. As a result, our current empirical evaluations are primarily constrained to offline public benchmarks, instead of online A/B tests. \textit{To bridge this gap and further validate our findings, we broadly welcome various forms of collaboration from the community and industry, including, but not limited to, the provision of real-world business scenarios and computational resources.} Interested researchers and practitioners are highly encouraged to reach out via Junmay.yin@connect.polyu.hk

 




\vfill

\end{document}